\documentclass[aps,prb,twocolumn]{revtex4-1}	
\usepackage{graphicx}
\usepackage{amsmath, amsfonts,amssymb}
\usepackage{color}

\newcommand{\be}{\begin{equation}}
\newcommand{\ee}{\end{equation}}

\newcommand{\bea}{\begin{eqnarray}}
\newcommand{\eea}{\end{eqnarray}}

\newcommand{\Tr}{{\rm Tr}}

\usepackage{outlines}

\usepackage{mathrsfs}

\renewcommand{\vec}[1]{{\bf #1}}

\renewcommand{\vec}[1]{{\bf #1}}

\begin{document}
\title{
 Topological Frequency Conversion in a Driven Dissipative Quantum Cavity}
\author{Frederik Nathan$^1$, Ivar Martin$^2$, Gil Refael$^3$}
\affiliation{$^1$Niels Bohr Institute, University of Copenhagen, 2100 Copenhagen, Denmark \\
$^2$Materials Science Division, Argonne National Laboratory, Illinois 60439, USA \\
$^3$Institute for Quantum Information and Matter, Caltech, Pasadena, California 91125, USA }
\begin{abstract}
Recent work (PRX {\bf 7}, 041008) shows that a spin coupled to two externally supplied circularly-polarized electromagnetic modes can effectuate a topological, quantized transfer of photons from one mode to the other.
Here we study the effect in the case when only one of the modes is externally provided, while the other is a dynamical quantum mechanical cavity mode. Focusing on the signatures and stability under experimentally accessible conditions, we show that the effect persists down to the  few-photon quantum limit and that it can  be used to generate highly entangled ``cat states'' of cavity and spin.
By tuning the strength of the external drive to a ``sweet spot", the quantized pumping can arise starting from an empty (zero photon) cavity state.
We also find that inclusion of external noise and dissipation does not suppress but rather {\it stabilizes} the conversion effect, even after multiple cavity modes are taken into account. 
\end{abstract}
\date{\today}
\maketitle

In recent years,  topology has played a major role in quantum physics, especially after the theoretical prediction~\cite{Thouless1982TKNN,Bernevig2006,Kane2005_1,Fu2006,Kitaev2009,Schnyder2010}
and subsequent discovery~\cite{Konig2007_1,Hsieh2008_1} of an extensive family of novel topological materials.
These phases of matter are characterized by highly nontrivial properties that lead to exotic but extremely robust phenomena with many possible applications.
More recent work has  shown that periodic driving can be used to induce analogous~\cite{
Yao2007,Oka2009,Inoue2010,Kitagawa2010,Lindner2011,Jotzu2014}, or entirely new\cite{
Jiang2011,Rudner2013,Nathan2015,Roy2016,vonKeyserlingk2016a,Potter2016,Else2016b,Khemani16,Choi16DTC,MonroeDTC,AFAI}  topological phases of matter in otherwise ordinary systems.   

Topological effects, however, are not limited to exotic phases of matter:
by replacing the spatial degrees of freedom of a topological insulator with other kinds of degrees of freedom (such as optical degrees of freedom, or tunable parameters) one can engineer systems that inherit topological features of the original models, but expose them in distinct and possibly useful ways.
The best-known example is perhaps  Thouless' adiabatic charge pump\cite{Thouless1983Pump}, which can be seen as a two-dimensional Chern insulator, where one of the spatial dimensions  is replaced with a tunable parameter; more recent examples of this idea include Refs.~\cite{Fu2006,EnergyPumpPaper,Peng2018,Weinberg2017,Crowley2018}, where, for instance, novel, energy pumping effects arise when the system is subject to external driving while the parameter is being tuned. 
Other examples of such analogies include optical\cite{Haldane2008,Khanikaev2012,Rechtsman2013,Gao2016,Mukherjee2017,Maczewsky2017,Barik2018} or acoustic\cite{Kane2013,Paulose2015,Susstrunk2015,Peng2016}  waveguide arrays, where nontrivial topology results in  protected, unidirectional modes of propagation, and systems where the electronic orbital degrees of freedom of a topological insulator are replaced with the angular momentum of light~\cite{Cardano2017}.

\begin{figure}[t]
\includegraphics[width=1\columnwidth]{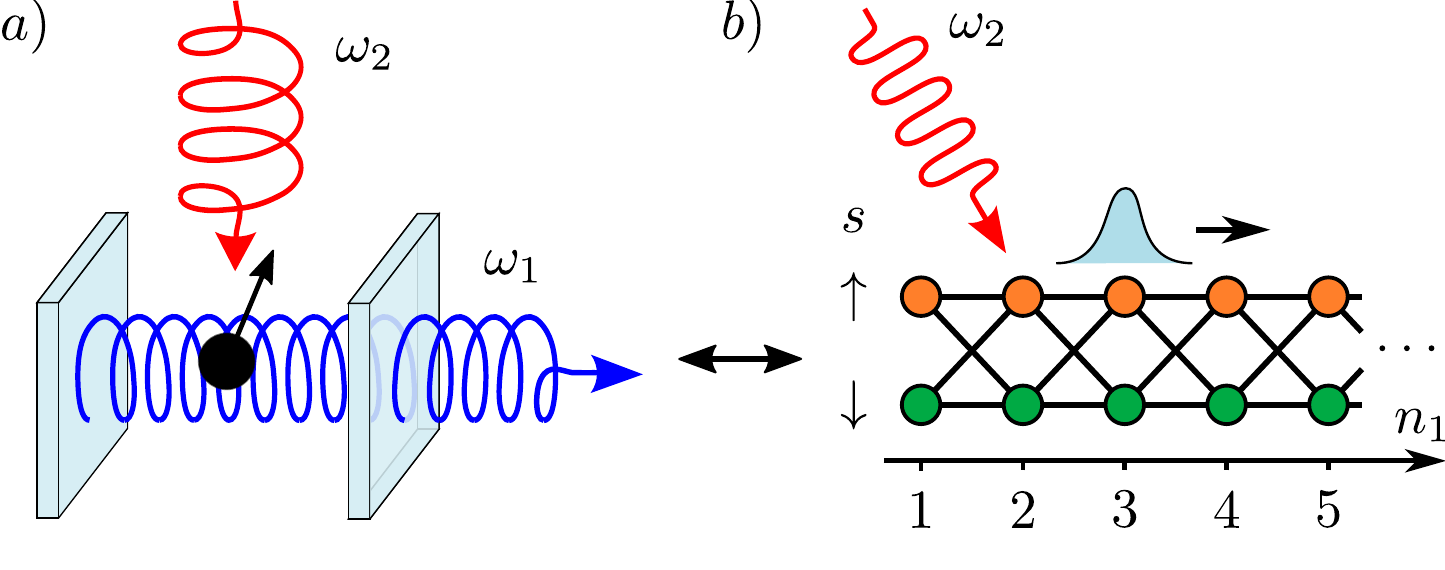}
\caption{ 
$(a)$ We study a model of a magnetic particle with angular moment $L$    (black dot, with arrow indicating magnetization) in a cavity, coupled to a circularly polarized cavity mode (blue) and a circularly-polarized driving mode (red). 
 Interpreting the particle's spin $s$ and the cavity photon number $n_1$  as an orbital and lattice degrees of freedom, respectively, the system is equivalent to a driven tight-binding model $(b)$. 
In suitable parameter regimes, the tight-binding model acts as a Thouless pump, and the magnetic particle effectuates a transfer of energy to the cavity   at the  universal rate $\omega_1 \omega_2  L /\pi$. 
}
\label{fig:FC:FrontPageFigure}
\end{figure}

In this work, we focus on a particular example of such a unconventional topological effect, which was first discussed in Ref.~\onlinecite{Martin2017}.  
We consider a magnetic particle with angular momentum $\ell \hbar /2$,
  coupled to two circularly-polarized electromagnetic modes, where one mode ($1$) is a cavity mode and the other ($2$) is externally driven (see Fig.~\ref{fig:FC:FrontPageFigure}a).
Interpreting the photon number of the cavity mode as a lattice degree of freedom, and the spin as an orbital degree of freedom, this setup is equivalent to a one-dimensional,  tight-binding model (see Fig.~\ref{fig:FC:FrontPageFigure}b).
In topologically-nontrivial parameter regimes, where the analogous lattice model acts as a Thouless pump, the number of photons in the cavity mode increases by the integer  $z$ for each cycle of the driving mode\cite{Martin2017}, and  
the spin affects a transfer of energy from the driving field to the cavity mode, at the topologically-quantized rate of $ \hbar \omega_1\omega_2 \ell /2\pi$, 
where $\omega_1$ and $\omega_2$ denote the angular frequencies of the two modes.
Thus, the cavity mode with frequency $\omega_1$ can be topologically pumped by driving the magnetic particle at the other frequency $\omega_2$. 
Importantly, $\omega_2$ does not need to be finely tuned, but can be set  arbitrarily as long as adiabaticity is respected.
This topological effect  opens up possibilities for optical amplification, or lasing at the frequency $\omega_1$, which is set by the properties of the cavity. 

The goal of this work is to explore the validity and robustness of the topological frequency pumping described above when taken to the quantum regime of realistic optical cavities and resonators. In particular, we consider an external periodic drive coupled through the spin with an optical cavity, which could be noisy and dissipative, as well as contain several modes.

First, we find that,  when treating the cavity modes as quantized  field,  the topological pumping effect persists all the way from the classical limit to the quantum mechanical  few-photon limit. 
Remarkably, the topological energy transfer arises spontaneously, even when the cavity initially holds zero photons. 

The system produces other surprises as well. The direction of transfer between the two modes is set by the alignment of the spin with the field.  When the system is isolated from the external environment, nontrivial ``cat'' states can therefore arise, in which the state of the spin is highly entangled with the cavity mode. We demonstrate the existence of these states numerically. 

To study extrinsic dissipation effects, we introduce a coupling between the cavity mode and the external electromagnetic environment (e.g., in the form of a semitransparent mirror in the cavity), and include the effects of extrinsic spin fluctuations in the model.  
The  resulting dynamics are simulated using a new Lindblad-form master equation which is is derived in a separate paper that will appear shortly~\cite{MarkovLindblad}.
The new master equation, referred to as the Markov-Lindblad  equation here, accurately describes open quantum systems where the correlation time of the external baths can neglected, and relies exclusively on the Markov-Born approximation. 
Crucially, the Markov-Lindblad equation is in the Lindblad form, and  can thus be integrated efficiently  with stochastic methods\cite{Molmer1993}.

Using this new method, we find that the quantized topological energy transfer persists even in the presence of cavity and spin dissipation. In fact, we find the  dissipation {\it stabilizes} the topological energy transfer:  dissipation in the spin's motion keeps the magnetic moment aligned with the field, while  cavity dissipation leads  to a steady state of the system, where the cavity emits a quantized number of  photons per driving period. 
These results demonstrate  that the topological energy transfer does not require coherence of the system's wave function.

The  stability of the topological energy transfer  in the presence of dissipation hints that the effect has a classical counterpart. 
We verify this intuition by demonstrating   the fully classical limit of the model supports a phase in which energy is transferred to the cavity mode at the universal rate
\be 
\dot E = \frac{\omega_1 \omega_2 L}{\pi},
\ee
where $L$ denotes the (macroscopic) angular moment of the magnetic particle. 
It is remarkable that topological effects, such as the effect discussed here, can  arise in relatively simple {\it classical} systems.  

Finally, we investigate the energy pumping effect when multiple cavity modes are included in the model, such as the higher harmonics of the fundamental cavity mode, as well as their time-reversed partners. 
Exploring the effects of these additional modes on the classical system, we find that the energy transfer effect is stable in certain parameter regimes  with a suitable, experimentally achievable engineering of the cavity. 

The rest of the manuscript is organized as follows:
in Sec.~\ref{sec:ModelIntroduction}, we  introduce the driven-cavity quantum model and demonstrate its topological energy transfer in the high-filling classical regime of the cavity.  
In Sec.~\ref{sec:FC:QuantumNoDissipation} we consider the behavior of the quantum model outside the ideal energy transfer region, and demonstrate the creation of photonic cat states, as well as energy transfer to an empty cavity. 
In Sec.~\ref{sec:FC:QuantumWithDissipation}, we introduce dissipation to the system, and use the new Markov-Lindblad  equation 
to study it.
Secs.~\ref{sec:Classical}-\ref{sec:MultipleModes} demonstrate that the effect persists in the classical version of the model, and in the presence of multiple modes.
 We conclude  by  discussing  the results of this paper, as well as possible experimental realizations  (e.g., with Weyl-semimetals, Yttrium-Iron Garnet (YIG) spheres, or mechanical gyroscopes)  in Sec.~\ref{sec:FC:Discussion}.

\section{The driven cavity model and topological energy transfer}
\label{sec:ModelIntroduction}

We begin by introducing the model that will be studied in this paper, and demonstrating how topological frequency conversion emerges in the model.

The system we consider consists of a magnetic particle with angular momentum $\vec L$, located  
 within a one-dimensional electromagnetic cavity with axis along the $x$-direction, as depicted in Fig.~\ref{fig:FC:FrontPageFigure}a.
In addition to the magnetic field from the cavity modes $\vec B_c$, the  particle is subject to the  field from 
   a circularly-polarized wave propagating  along the $y$-direction $\vec B_d(t)= (B_d \sin\Omega t,0,  -B_d \cos\Omega t)$, as well as an  static (Zeeman) magnetic field $\vec B_m=(0,0,B_m)$ applied along the $z$ direction. 

In general, the cavity field $\vec B_{\rm c}$ is a  superposition of multiple distinct modes from the discrete frequency spectrum of the cavity. 
Topological energy transfer arises when the cavity field $\vec B_c$ is dominated by a single  circularly-polarized  mode:
$\vec B_c= (0, B_c\sin\phi, -B_c\cos\phi)$, where $B_c$ and $\phi$ denote the amplitude and phase of the dominant mode. 
Ignoring the effects of the additional modes,  $B_c$ and $\phi$, along with the angular moment of the magnetic particle $\vec L$  constitute the dynamical variables of the system.
In Sec.~\ref{sec:MultipleModes} we discuss the effects of taking the additional modes into account. 

Including the energy of the dominant cavity mode, the Hamiltonian of the combined particle-cavity system reads 
\begin{align}
H(t)=  \frac{V}{\mu}B_c^2  - g \left[\vec B_{\rm c}+\vec B_{m}+ \vec B_{ d}(t)\right] \cdot \vec L. 
\label{eq:Hamiltonian0}
\end{align}
where $V$ and $\mu$ respectively denote the cavity's  volume and magnetic permeability,  while $g$  is the (isotropic)  gyromagnetic ratio of the magnetic particle.
The quantum Hamiltonian  of the  system $ H(t)$ is obtained through canonical quantization of Eq.~\eqref{eq:Hamiltonian0}. 
In particular, the cavity mode is quantized by replacing $B_c e^{-i\phi} $ with $ 
B_0  \hat a$, 
where  $ \hat a $ denotes  the photon annihilation operator of the cavity mode, and $B_0 \equiv \sqrt{{\omega \mu_0 \hbar }/V}$, where $\omega$ is the cavity mode's frequency. 
Physically, $B_0$ gives the magnetic field amplitude   corresponding to a single photon in the cavity. 
This results in the quantum Hamiltonian
\be 
 H(t) = hbar  \omega \hat  n  -   g \,  \hat{\vec{  B}}(t)\cdot  \hat{\vec{ L}}.
\label{eq:QuantumHamiltonian}
\ee
Here $\hat n \equiv \hat a^\dagger \hat a$ is the photon number operator, ${\vec L}$ denotes the quantum mechanical angular momentum operator of the magnetic particle, and 
 \begin{eqnarray}
 \hat  B_x(t) &=&   B_d \sin(\Omega t),
 \label{eq:hform1} \\
   \hat B_y (t) &=& B_0   \frac{ \hat a- \hat a^\dagger}{2i}, \\
 \hat B_z (t) &=&   B_m-  B_d \cos(\Omega t)- B_0\frac{\hat a+\hat a^\dagger}{2}.
\label{eq:hform3}
\end{eqnarray}
Here we ignored the constant shift  $\hbar \omega /2$ to  the Hamiltonian $ H(t)$ from the cavity's vacuum energy. 

The Hilbert space on which $ H(t)$ acts is spanned by the basis $\{|n,s\rangle\}$, where $n\ge 0$ is the number of photons in the 
cavity mode, and $s$ is the angular momentum of the magnetic particle along the $z$-axis, in units of $\hbar$, satisfiying $-\ell/2\le s \le \ell/2$.
The photon annihilation operator $ a$ and the angular momentum operator $\vec L$ respectively act on the photon and spin degrees of freedom of the states. 
In particular, $ \hat a|n,s\rangle = \sqrt{n}|n-1,s\rangle$. 

One may view the system above as a  one-dimensional, semi-infinite tight-binding model, which we refer to as the photon lattice in the following. 
The photon number $n$ can be interpreted as a lattice index, and the spin index $s$ as an orbital index. 
 In this picture, the state of the system is represented by the wave function of a single particle in the photon lattice, and the Hamiltonian $ H(t)$ couples neighboring sites through the photon creation and annihilation operators $\hat a , \hat a^\dagger$ (see Fig.~\ref{fig:FC:FrontPageFigure}b), 
\subsection{Topological frequency conversion}

\label{sec:TopologicalFrequencyConversion}
\begin{figure}
\includegraphics[width=\columnwidth]{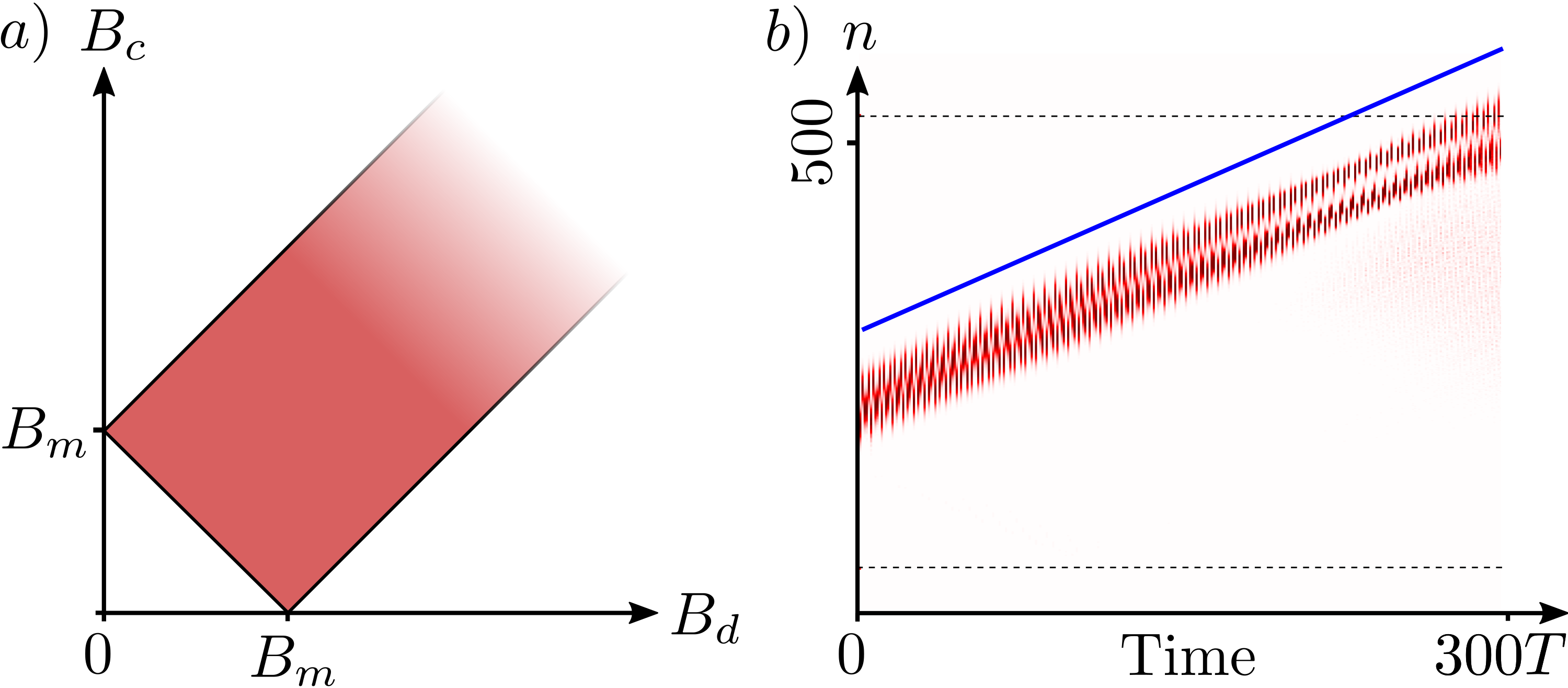}
\caption{
$a)$  Region of parameter space which supports topological frequency conversion (red). 
Here  $B_d$ and $B_c$ respectively denote the amplitudes of the driving and cavity modes, while $B_m$ indicates the strength of the Zeeman field. 
$b)$ Numerically obtained  wave function of the system (absolute squared) as a function of photon number $n$ and time, in the topological regime -- see main text for further details.
The  boundaries of the topological regime (corresponding to the boundaries of the highlighted region in panel $a$) are indicated by   dashed lines, and the quantized pumping rate is indicated by the slope of the blue line.
 }
\label{fig:PhaseDiagram}
\end{figure}

To demonstrate how the topological energy transfer effect emerges,  
we consider a state $|\Psi\rangle$  whose support  is confined to photon numbers $n$ in some relatively narrow region $\mathcal R$ centered around $n=n_c$, and of width 
 $\Delta n \ll n_c$. 
Using that $\sqrt{n}=\sqrt{n_c}(1+\mathcal O(\Delta n/n_c))$ when $n$ is in the interval $\mathcal R$, we find the following simple expression for the action of photon annihilation operator $ a$ on the state of the system $|\Psi\rangle$: 
\be 
\hat a|\Psi\rangle = \frac{B_c}{B_0} \hat T 
|\Psi\rangle  + \mathcal O \left(\frac{\Delta n  B_0}{B_c}\right).
\label{eq:AnnihilationOperatorApproximateExpression}
\ee
Here  $B_c\equiv  B_0\sqrt{n_c}$ is the   amplitude of the cavity field consistent with photon number $n_c$, and $ \hat T$ acts as the translation operator in the photon lattice:
\[
\hat T= \sum\limits_{n=0}^\infty \sum\limits_l |n-1,l \rangle \langle n,l|  .
\]
In the limit we study,  $\Delta n \ll (B_c/B_0)^2$.
Hence the correction in Eq.~\eqref{eq:AnnihilationOperatorApproximateExpression} is negligible, and can be ignored. 
Using Eq.~\eqref{eq:AnnihilationOperatorApproximateExpression} in Eqs.~\eqref{eq:QuantumHamiltonian}-\eqref{eq:hform3},
we thus find that the state $|\Psi\rangle$ evolves with the Hamiltonian 
\be 
 H_{\rm eff} =  H_T(t)+  \omega  n ,
\label{eq:HeffDef} 
 \ee
 where  the ``Thouless Hamiltonian'' $ H_T$ is given by    
\begin{align*}
 H_{T}(t)  =&  -   \frac{g B_c}{2i}(\hat T - \hat T^\dagger)  \hat L_x - g B_d \sin(\Omega t )\hat  L_y  \\ 
&-g \left(B_m-\frac{B_c}{2}( \hat T+ \hat T^\dagger)-B_d\cos(\Omega t)\right)\hat  L_z   
\end{align*}
Ignoring boundary effects, we note that the ``Thouless Hamiltonian'' $ H_T(t)$ is invariant under translations in photon space.
Hence, the effective Hamiltonian $ H_{\rm eff}(t)$   describes a one-dimensional translationally-invariant lattice model with a constant longitudinal electric field $\omega$.  

The effective Hamiltonian $ H_{\rm eff}(t)$ describes an adiabatic ``charge pump'' in the  photon lattice, with {quantized} average group velocity of wave-packets.
To establish this result, we first consider the translationally-invariant Hamiltonian $ H_T(t)$ which forms the first part of $ H_{\rm eff}(t)$. 
The Bloch Hamiltonian associated with $ H_T(t)$ is given by 
\be 
 H_T(k,t) =  -g \vec B(k,t) \cdot  {\vec L}.
\ee
Here $k$ denotes the dimensionless crystal momentum in the photon lattice,   the operator $\hat{\vec L}$ acts on the $(\ell+1)$-dimensional  orbital space of the tight-binding model, and 
\be 
\vec B(k,t) =
 \left(\begin{array}{c}
B_d\sin(\Omega t) 
 \\  B_c \sin(k) \\
  B_m- B_c \cos(k)-B_d \cos(\Omega t ))
\end{array}\right).
\label{eq:EffectiveFieldDef}
\ee

To analyze the dynamics of the system,  we  decompose the  initial state  of the system $|\Psi_0\rangle$  in a superposition of crystal momentum eigenstates in the photon lattice: 
\be
|\Psi_0\rangle = \int_0^{2 \pi}\!\! dk\, f(k) |\psi_0(k)\rangle\otimes|k\rangle,
\ee
where  the normalized state $|\psi_0(k)\rangle$ lives in the orbital space of the system, while $|k\rangle$ denotes the state in the ``lattice space'' with crystal momentum $k$, such that $ T|k\rangle = e^{i k }|k\rangle$. 
Here  $f(k)$ is a positive weight factor satisfying $\int_0^{2\pi}dk |f^2(k)|=1$.
Using that $\hat n |k\rangle= i \partial _k |k\rangle$, 
 one can verify  that the time-evolution of the system 
is given by 
\be 
|\Psi(t)\rangle = \int_0^{2\pi}\!\! dk \,  f(k) |\psi(k, t) \rangle\otimes |k-\omega t\rangle .
\label{eq:SchrodingerEquationSolution}
\ee
Here $|\psi(k,t)\rangle$ solves the Schr\"{o}dinger  equation
\be 
\partial _t |\psi(k,t)\rangle = -i  H_T (k-\omega t,t)|\psi(k,t)\rangle
\label{eq:PsiSchrodingerEquation}
\ee
with the initial condition $|\psi(k,0)\rangle = |\psi_0(k)\rangle$.

The quantization of angular momentum implies that $ H_T(k,t)$ 
has $\ell+1$ energy bands. 
The bands of $ H_T(k,t)$ are evenly spaced, with the  gap width (as a function of $k$ and $t$)  given by $\Delta E(k,t)= g |\vec B(k, t)|$. 
As can be seen from Eq.~\eqref{eq:EffectiveFieldDef}, the field $\vec B(k,t)$ can be  obtained by replacing the operator  $B_0  a$ with $B_c e^{i k}$ in Eqs~\eqref{eq:hform1}-\eqref{eq:hform3}. 
In this way,  $ H_T(k,t)$ can be seen as the Hamiltonian acting on the  magnetic particle at time $t$, when the cavity mode is classical and has   amplitude $B_c$ and phase $-k$. 
Each band of $ H_T (k,t)$ thus corresponds to a distinct projection of the spin onto the  resulting classical magnetic field $\vec B(k,t)$ at time $t$. 

The quantized energy transfer occurs in the near-adiabatic limit, when  
  the instantaneous energy gap (corresponding to the spin's  precession frequency)
   $\Delta E(k,t)$ is large compared to  the frequencies $\Omega$ and $\omega$ for all $k,\,t$. 
One can verify that  $\Delta E(k,t)$ takes minimal value $\Delta E_{\rm min}
 = \min g |B_c  - B_{\pm}|$, where $B_{\pm} \equiv|B_m\pm B_d|$.
Thus, for topological energy transfer to emerge, we require $|B_c - B_{\pm}|\gg \Omega / g$.
%

To show how the energy pumping arises, consider the system initialized in a wave-packet constructed from the $m$th band of $ H_T(k,t)$: $|\psi_0(k)\rangle =| \phi_m(k,0)\rangle$, where $|\phi_m(k,t)\rangle$ denotes the $m$th eigenstate of $ H_T(k,t)$.
Since $\omega$ and $ \Omega$ are much smaller than the level spacing of $ H_T(k,t)$ 
the  state $|\psi(k,t)\rangle$  in Eqs.~\eqref{eq:SchrodingerEquationSolution}-\eqref{eq:PsiSchrodingerEquation} evolves according to the adiabatic theorem: $|\psi(k,t)\rangle = e^{-i \theta(k,t)}|\phi_m(k-\omega t , t)\rangle$, where $\theta(k,t) = \int_0^t dt E_m(k-\omega t)$ denotes the dynamical phase.  

Having established how $|\psi(k,t)\rangle$ evolves, we now compute the number of photons in the system-bath state. 
Using that $ \hat n |k\rangle= i \partial _k|k\rangle $   in Eq.~\eqref{eq:SchrodingerEquationSolution}, and by integrating by parts, we find, after using  $\int_0^{2\pi} dk\partial _k|f^2(k)|=0$,
\be 
\langle n(t) \rangle  = i  \int_0^{2\pi} \!\! dk\, \langle \psi(k,t)|\partial _k  |\psi(k,t)\rangle \, |f^2(k)| .
\ee
Substituting in $|\psi(k,t)\rangle = e^{-i \theta(k,t)}|\phi_m(k-\omega t , t)\rangle$, and taking the time-derivative, we find 
\be 
\partial _t \langle n\rangle = \int \!dk  \, |f^2 (k)|v_m(k-\omega t,t),
\ee
where the ``velocity'' in the photon lattice $v_m(k,t)$ is given by 
 \be 
 v_m(k,t) =  \Omega_m(k,t) + \frac{\partial E_m(k,t)}{\partial k}.
 \ee
 Here $\Omega_m(k,t)$ denotes the Berry curvature of band $m$ and is given by $\Omega_m(k,t) \equiv i(\partial _k \langle \phi_n(k,t)|\partial _t |\phi_n(k,t)\rangle - \partial _t \langle \phi_n(k,t)|\partial _k |\phi_n(k,t)\rangle)$.

As a next crucial step, we note that the velocity $v_m(k,t)$ is periodic in both of its arguments, and lives on the torus $\mathbb T \equiv [0,2\pi]\otimes [0,T]$, where $T\equiv \frac{2\pi}{\Omega}$ denotes the period of the driving mode. 
Hence, if the frequencies $\omega $ and $\Omega$ are incommensurate, averaging $v_m(k+\omega t,\Omega t)$ over time amounts to averaging the function $v_m(k,t)$ over the entire torus $\mathbb T$. 
Using that the integral $\int_0^{2\pi} dk \frac{\partial E_m}{\partial k}$ vanishes, and letting $\overline{F}$ denote the time-average of the function $F(t)$, we find
\be 
\overline {\partial _t\langle {  n}\rangle}  =\frac{ C_m }{T}.
\label{eq:dNdTQuantization}
\ee
Here $C_m  \equiv \frac{1}{2\pi   } \int dk dt\, \Omega_m(k,t) $ is the Chern number of the $m$th band of $
 H_T(k,t)$, and takes integer values.

It is well-established that the $n$th-lowest band of $H_T(k,t)$  has Chern number $C_m=\ell-2(m-1)$~\cite{ThoulessPump}   when  $B_c$ is between $B_-$ and $B_+$.
In this case, the Chern number of the lowest band (corresponding to the spin being aligned with the net instantaneous field) is given by $\ell $. 
The region of parameter space where $B_c\in [B_-,B_+]$ is indicated in Fig.~\ref{fig:PhaseDiagram}a.

For the derivation of Eq.~\eqref{eq:dNdTQuantization}  to be valid, the frequencies $\omega$ and $\Omega$ 
must be small compared to the gap $\Delta E_{\rm min} = g \min|B_c - B_{\pm}|$. 
Thus,  for the topological energy transfer to take place 
the cavity amplitude $B_c$ 
must fall within $[B_-,B_+]$ (the interior of the red region in Fig.~\ref{fig:PhaseDiagram}a), and be more than a distance $\sim \Omega /g$ away from the region's boundaries $B_-$ and $B_+$. 

When the conditions above are met, and the spin is aligned with the instantaneous field, we conclude from the discussion above that 
 the   number of cavity photons on average increases at the quantized rate 
\be 
\overline{\partial _t \langle  n \rangle}  = \frac{\ell }{T}. 
\label{eq:DotNQuantizationResult}
\ee
where $T \equiv \frac{2\pi}{\Omega}$ denotes the period of the driving mode. 
Using that the cavity's field energy is given by $E_c = \hbar \omega \langle n\rangle$, and that $\ell = 2 L/\hbar$, where $L$ is the magnetic particle's angular moment, yields that the energy is transferred to the cavity at the universal rate 
\be 
\dot E = \frac{\omega  \Omega L }{\pi}.
\ee
Interestingly, as we will demonstrate in Sec.~\ref{sec:Classical}, the above result is not only a quantum effect, but persists in the macroscopic classical limit.

\subsection{Numerical simulations}
\label{sec:NonDissNumerics}

We verified the conclusions of the previous subsection numerically in a model  with a spin-$1/2$ magnetic particle  (Fig.~\ref{fig:PhaseDiagram}b). 
Using the quantities $\Omega$ and $B_0$ to set the scales in the simulation [see text above Eq.~\eqref{eq:QuantumHamiltonian}], 
we chose  the parameters   $ B_m=15B_0$, $B_d=8B_0$, $g = 2 \Omega/B_0$, and $\omega = \Omega/\varphi$,  where the irrational number  $\varphi $ is given  by $\varphi= (\sqrt{5}-1)/2$ 
 (recall that the quantized energy transfer requires the two modes' frequencies to be incommensurate). 
From the discussion in the end of the above subsection, the topological pumping 
 arises when $\sqrt{n}  = B_c/B_0  $ is in the interval  between $7$ and $23$, where $n$ denotes the photon number in the cavity. 
 Moreover $\sqrt{n}$ should not be  closer than $\mathcal O(\Omega/g B_0) \sim  O(1)$ from either of the boundaries. 
 The two boundaries correspond  to $ n=7^2 = 49$ and $n=23^2 = 529$ photons.

 The system was  initialized in a direct product of a coherent cavity  state and a spin state. 
The coherent cavity state was centred around $200$ photons and had  phase zero, while the spin was initialized in the state $|\!\! \downarrow\rangle$, corresponding to alignment with the net resulting field at time $t=0$. 
The system was then evolved by direct time-evolution, with the first 600 photon states included in the simulation.
In Fig.~\ref{fig:PhaseDiagram}b, we plot the resulting evolution of the system's wave function (absolute square) as function of photon number $n$  and time. 
The dashed horizontal lines indicate the phase boundaries $n=49$ and $n=529$, while the solid blue line indicates the slope corresponding to the photon number $n$ increasing by $1$ per driving period. 
As can be seen in the figure, the photon number increases with the quantized rate $\frac{1}{T}$ (indicated by blue line), as expected from Eq.~\eqref{eq:DotNQuantizationResult}.

\section{Unitary 
dynamics}

\label{sec:FC:QuantumNoDissipation}
\begin{figure}
\includegraphics[width=1\columnwidth]{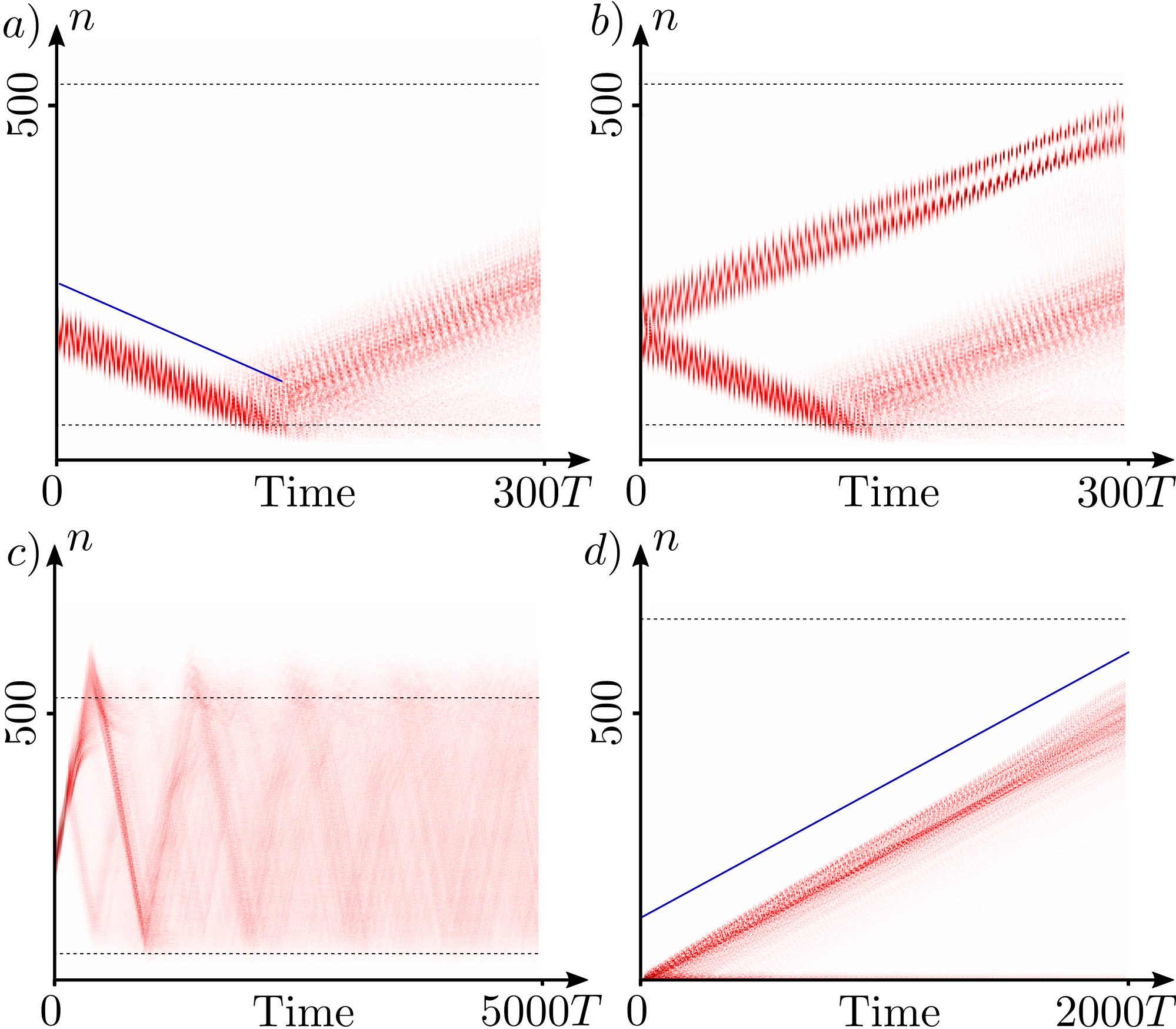}
\caption{ Unitary evolution of the system's wavefunction (see main text for further details).
$a)$ Evolution of the system's wavefunction for the model depicted in Fig.~\ref{fig:PhaseDiagram}b, when the  spin is initially anti-aligned with the net magnetic  field.
$b)$ Evolution of the system's wavefunction for the same setup as in panel $a)$, when the spin is initially perpendicular to the net field. 
$c)$ Evolution of the system in Fig.~\ref{fig:PhaseDiagram}b over $5000$ periods. 
$d)$ Evolution of the model at the ``sweet spot'' in the phase diagram (see Fig.~\ref{fig:PhaseDiagram}a), $B_d=B_m = 10B_0$, when the cavity is initially empty.
Black dashed lines indicate phase boundaries, and  blue dashed lines indicate the theoretically expected rate of  $\langle \dot n \rangle = \frac{1}{T}$. 
} 
\label{fig:QuantumSimulationsNoDissipation}
\end{figure}
Let us next explore the pumping effect beyond the simplest regime of the previous section. We first show that the system can create photonic cat states when the magnetic particle is initialized in a superposition of instantaneous eigenstates. We then consider how the system behaves near the  parameter-space boundaries of validity (outlined in Fig. \ref{fig:PhaseDiagram}a). Perhaps the most intriguing result we encounter is that the pumping effect can populate a cavity mode even in the extreme quantum limit, when the cavity is initially empty, well beyond the regime of applicability of the analysis of Sec.~\ref{sec:ModelIntroduction}.


\subsection{Producing cavity cat states}

The result in Eq.~\eqref{eq:DotNQuantizationResult} intimately links the state of the spin (i.e. the ``orbital'' of the wave packet in the photon lattice) with the state of the cavity (the ``lattice'' part of the wave function). 
In particular, Eq.~\eqref{eq:DotNQuantizationResult} suggests that the photon number's rate of change  is sensitive to the alignment of the spin with the net resulting field. 
In particular, while alignment of the spin with the field results in an increase of the photon number (see Fig.~\ref{fig:PhaseDiagram}b),  the photon number should {\it decrease} if the spin is {anti-aligned} with the net instantaneous field. 

This conclusion is verified numerically in Fig.~\ref{fig:QuantumSimulationsNoDissipation}. 
Here we  study the same setup as in the end of Sec.~\ref{sec:TopologicalFrequencyConversion},  except that the spin is initialized in the state $|\!\!\uparrow\rangle$, corresponding to anti-alignment with the initial field. 
As can be seen in Fig.~\ref{fig:QuantumSimulationsNoDissipation}b, the photon number in this case decreases at the average rate $-1/T$ (solid blue line), as predicted by Eq.~\eqref{eq:DotNQuantizationResult}. 

Note that any state of the system can be decomposed into wave-packets constructed from the distinct bands of $ H_T (k,t)$. 
These different components of the state experience different rates of photon pumping, according to Eq.~\eqref{eq:dNdTQuantization}. 
Thus, when evolved from an arbitrary state with a relatively well-defined photon number, the state of the system should split into a ``cat'' state where the  photon number in the evolved state is highly entangled with the state of the spin. 
 This effect is demonstrated in Fig.~\ref{fig:QuantumSimulationsNoDissipation}b. 
 Here we consider the exact same setup as in Fig.~\ref{fig:PhaseDiagram}b and Fig.~\ref{fig:QuantumSimulationsNoDissipation}a, but initially polarize the spin along the $x$-axis, such that the state of the spin is initially given by $\frac{1}{\sqrt{2}}(|\!\!\uparrow\rangle + |\!\!\downarrow\rangle)$. 
 The wave-packet of the system in the photon lattice is thus in an equal superposition of the two bands of $ H_T(t)$. 
 As a result, the two components of the system's state (the aligned and antialigned) evolve in different directions in the photon lattice. 
 For the resulting final state, the photon number is highly entangled with the spin of the magnetic particle. 
 
\subsection{Behavior near boundaries of the topological region}
\label{sec:BoundaryBehaviour}
We now consider the behavior of the system near the two phase boundaries at $B_c=B_{-}$ and $B_c=B_+$. 
Suppose the spin is initially anti-aligned with the field, 
 as in Fig.~\ref{fig:QuantumSimulationsNoDissipation}a.
In this  case, the photon number will constantly decrease until the wave-packet of the system approaches one of the phase boundaries. 
The photon number cannot decrease below the phase boundary, since the average group velocity of wave-packets is zero in this region.
Instead, the wave-packet will be reflected at the phase boundary:
as the wave function approaches the boundary, the gap $\Delta E_{\rm min}$ of $ H_T(k,t)$ will  at some point become comparable to the driving frequencies $\omega$ or $\Omega$. 
In this case, the assumption of adiabaticity made in Sec.~\ref{sec:TopologicalFrequencyConversion} breaks down. 
Due to the small size of the gap near the boundary, the system's wavefunction  will at this point undergo partial Landau-Zener tunneling to other bands of $ H_T(k,t)$ (corresponding to the spin changing its alignment with the instantaneous field). 
Eventually, parts of the wave functions will tunnel to a band where the photon number {\it increases}, transporting the wave function {away} from the boundary.
This mechanism allows the system to escape from the boundary, by being pumped to another band. 

The process described above (referred to as a ``Landau-Zener reflection'' in the following) can be observed in Fig.~\ref{fig:QuantumSimulationsNoDissipation}b. 
Here, after approximately 120 driving periods,  the system's wave function deflects from a trajectory with decreasing photon number to a trajectory with increasing photon number.  
The deflection clearly occurs when the photon number  reaches the topological phase boundary at $49$, which were predicted in  Sec.~\ref{sec:TopologicalFrequencyConversion}. 

A Landau-Zener reflection may take several driving periods to complete, since the system only has a finite probability of Landau-Zener tunnelling to another band   each driving period. 
As a result, the system's wavefunction gets ``smeared out in time'' during each Landau-Zener reflection. 
If the system evolves undisturbed over long time intervals, multiple Landau Zener reflections may occur. 
At each reflection, the wave function gets increasingly smeared out in time.  
In Fig~\ref{fig:QuantumSimulationsNoDissipation}c, we evolve the wave function for the system in Fig.~\ref{fig:PhaseDiagram}b over $5000$ driving periods, with $700$ photon states included in the simulation (such that the upper phase boundary at 529 photons is far below the photon number cutoff).
We observe the state of the system undergoing approximately $9$ ``Landau-Zener reflections'', with the wave function getting increasingly smeared out at each reflection. 
Note that 
the wave-function clearly remains confined  between the topological boundaries at $\sim 530$ and $\sim 50$ photons at all times, despite of being  ``smeared out''.
The clearly visible boundaries at photon number $530$ and $49$ are in excellent agreement with the theoretical predictions of Sec.~\ref{sec:TopologicalFrequencyConversion}.


\subsection{Topological pumping of initially empty cavity}
\label{sec:EmptyCavity}

We finally demonstrate that the energy transfer may  arise even when the cavity is initially empty. 
Specifically, consider the special point in the phase diagram where $B_d = B_m$. 
In this case, the lower phase boundary is located at $B_-  =0$, and one would naively expect the topological pumping effect in Eq.~\eqref{eq:DotNQuantizationResult} to arise even when the cavity holds zero photons ($B_c =0$, see Fig.~\ref{fig:PhaseDiagram}a). 
Note that the analysis  in Sec.~\ref{sec:TopologicalFrequencyConversion} does not apply in this limit, since the assumptions of  large photon number and adiabaticity are not valid.
In spite of this, the  naive expectation above holds up in numerical simulations:
we considered the system with the parameters $B_d = B_m = 10B_0$, and initialized the cavity in the vacuum state ($0$ photons), and aligned the spin along the $x$-axis.
Fig.~\ref{fig:QuantumSimulationsNoDissipation}d shows the resulting evolution of the state's wave function. 
As can be seen, the photon number of the cavity increases from zero at the quantized rate of one photon per driving period. 

The above results show that the topological energy transfer can arise  beyond regimes where the analysis in Sec.~\ref{sec:TopologicalFrequencyConversion} is valid. 
In particular, the pumping effect can arise spontaneously in an empty cavity, without requiring the cavity to be populated initially. 
Thus, importantly, the effect described in this paper (and Ref.~\onlinecite{Martin2017}) can in principle  be used generate photons with  desired frequency $\omega$, using an external pump with more readily available frequency.

\section{Floquet-Lindblad Equation and Dissipative Dynamics}
\label{sec:FC:QuantumWithDissipation}

We now discuss the behaviour of the driven cavity-spin system when dissipation is introduced. 
We consider two sources of dissipation: dissipation from the cavity and spin relaxation. 
The cavity dissipation can for instance arise from  a partially-transparent mirror in the cavity that couples cavity  photons to the outside electromagnetic continuum.
Spin relaxation is introduced to model a more realistic setting, where the magnetic energy of the spin can dissipate, for instance, due to spin-lattice relaxation.  

Below, we find that  the energy transfer effect persists and in fact can be enhanced by dissipation, and that the two dissipation mechanisms lead to new interesting effects. 
In particular, the introduction of cavity dissipation allows the system to reach a  ``lasing'' steady  state where 
the energy pumped into the cavity mode via frequency conversion (at the universal rate ${L \Omega  \omega }/{\pi}$) is exactly compensated by the cavity leakage loss.
On the other hand, spin dissipation  leads to  a stabilization of the energy pumping  beyond the near-adiabatic  regime.
Indeed, the quantization of energy transfer depends on the degree of alignment of the spin with the field, which  in the absence of dissipation is ensured by adiabaticity condition $g |\vec B|  \gg  \Omega $.
Spin dissipation increases the tendency of the spin to align with the field, and thus stabilizes the quantized energy transfer beyond the frequency ranges allowed by the adiabaticity condition. 

\subsection{Markov-Lindblad equation}
Following the standard approach for dissipative quantum systems~\cite{GardinerZoller}, we include  cavity dissipation in the model by connecting the cavity field, $\frac{B_0}{2}( a +  a^\dagger)$ to an external bosonic bath, with coupling strength $\gamma_c$. 
Here $\gamma_c$  is  for example set by  the transmission coefficient of a partially-transparent mirror in the cavity. 
Likewise, spin dissipation is modeled by coupling each of the spin's components $ L_x,  L_y, L_z$  to individual bosonic baths with a coupling strength $\gamma_s$ (assumed to be the same for all components). 
For both the cavity and spin degrees of freedom, we take the associated baths to be Ohmic, meaning that the spectral function $S(\omega)$ of each bath is linear in $\omega$~\cite{QuantumDissipativeSystems}: $S_{c,s}(\omega)=S(\omega)\equiv \frac{\omega}{\omega_0}$, where $\omega_0$ is some fixed energy scale.
Absorbing all variable parameters into the coupling strengths $\gamma_c$, $\gamma_s$, we set  $\omega_0=1/T$. 

With the four bosonic baths included in the model, the dynamics of the system can in principle be obtained by computing the time-evolution generated by the full system-bath Hamiltonian. 
From the resulting  system-bath state $|\Psi_{SB}(t)\rangle$, we can   compute any system observable using the reduced density matrix $ \rho (t)\equiv \Tr_B |\Psi_{SB}(t)\rangle\langle\Psi_{SB}(t)|$, where $\Tr_B$ traces out all  bath degrees of freedom.

Often,  the physical baths are found to be short-time correlated compared to  relevant time scales of the system, meaning that the system's density matrix  $ \rho$ at time $t+dt$ is fully determined by the density matrix at time $t$.
When this is the case, the evolution of $ \rho(t)$  can be described by a linear first order differential equation, referred to as a master equation.

A commonly used approach\cite{GardinerZoller} for obtaining such a master equation  is to employ the Markov-Born approximation, which results in the Redfield equation.
However, the Redfield equation  does not preserve the complete positivity of the density matrix, and  can be numerically expensive to solve for large systems. 
In order to simulate the system efficiently, we therefore use a new, alternative master equation, referred to as the Markov-Lindblad  equation here. 
The Markov-Lindblad equation is systematically derived from the microscopic model of the system-bath setup, using  only  the Markov-Born approximation (i.e.  that the correlation time of the bath can be ignored). 
Importantly, in contrast to the Redfield equation, the Markov-Lindblad  equation is by construction   in the  Lindblad form.
Hence the Markov-Lindblad equation preserves the unit trace and complete positivity of the density matrix, and can be efficiently be integrated with stochastic methods~\cite{Molmer1993}. 

A derivation of the Markov-Lindblad equation is  beyond the scope of this work,  and  will therefore be presented  in separate  work which is to appear shortly~\cite{MarkovLindblad}. 
%
%
In this paper, we  use the result of this work, which states that the reduced density matrix of the system evolves according to the following master equation:
\be 
\partial _t  \rho = -i[ H, \rho] + \sum_\nu\left( L_\nu \rho  L^\dagger_\nu - \frac{1}{2}\{ L_\nu L_\nu,\rho\}\right).
\label{eq:RhoMasterEquation1}
\ee
Here $ H(t) $ denotes the Hamiltonian of the system [Eq.~\eqref{eq:QuantumHamiltonian}], and the $\nu$-sum runs over the four different channels of dissipation (we suppressed the time-dependence above for brevity). 
The so-called jump operators $\{ C_\nu(t)\}$ are time-dependent with the same periodicity as the Hamiltonian, and are defined from the quasienergies $\{\varepsilon _a\}$ and the time-periodic Floquet states $|\phi_a(t)\rangle$\cite{FN:FloquetStates} %
 of the time-periodic Hamiltonian $H(t)$ as follows:
\be
 L_\nu(t) = \!\!\sum_{a,b,z}\!\! |\phi_a(t)\rangle\langle \phi_b(t)|e^{-iz \Omega t}L^{ab}_\nu[z].
\label{eq:L(t)Def}
\ee
Here the coefficient $ L^{ab}_\nu[z]$ is given by $L^{ab}_{\nu}[z]=\sqrt{2\pi \gamma_\nu J_\nu(\varepsilon _{b}-\varepsilon _a +z \Omega )} \mathcal O_{\nu}^{ab}[z]$, where 
\be 
\mathcal O_{\nu}^{ab}[z] = \frac{1}{T} \int dt \langle \phi_a(t)|\hat{\mathcal {O}}_\nu|\phi_b(t)\rangle e^{i\Omega zt}.
\label{eq:RhoMasterEquation3}
\ee
and where  $\hat{\mathcal {O}}_\nu$ is the operator connected to bath $\nu$ (i.e. either $\frac{B_0}{2} ( a+ a^\dagger)$, or $ L_{x,y,z}$).
Finally,  
$J_\nu(\omega) = S_\nu(|\omega|) \left[\theta(\omega) +n_\nu(|\omega|)\right]$, where $S_\nu(\omega)$  is the spectral function of bath $\nu$, $\theta(\omega)$ denotes the Heaviside function, and $n_\nu(\omega)$ is the thermal expectation value in  bath $\nu$ of the photon number  at frequency $\omega$.
In our case, we set the bath temperatures to zero, so $J_\nu(\omega)= \omega \theta(\omega)/\omega_0$.

\subsection{Dissipative dynamics} 
\label{sec:NumericsWithDissipation}
\begin{figure}
\center{
\includegraphics[width=1\columnwidth]{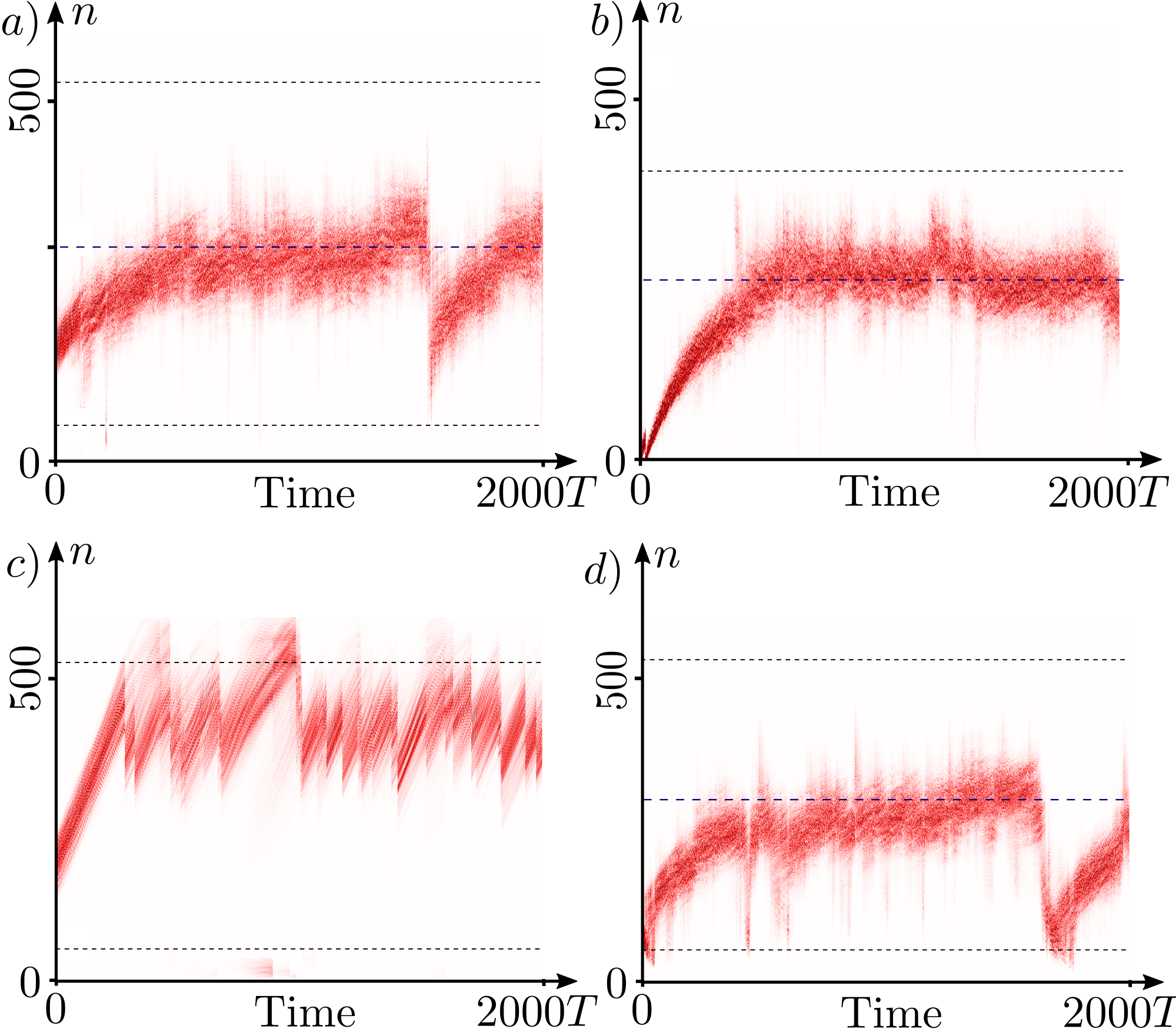}
\caption{
Evolution of the system in the presence  of noise and dissipation. 
Each panel plots a randomly picked example of the system's wave function (absolute squared), as a function of photon number $n$ and time, obtained by stochastic integration of  the Floquet-Lindblad master equation in Eqs.~\eqref{eq:RhoMasterEquation1}-\eqref{eq:RhoMasterEquation3}.
Parameters for the respective panels are given in Sec.~\ref{sec:NumericsWithDissipation}.
Black dashed lines indicate the topological phase boundaries, and  blue dashed lines indicate the expected steady state photon number consistent with a quantized emission rate (see Sec.~\ref{sec:NumericsWithDissipation}). 
$a)$ Example of the wavefunction when  spin and cavity dissipation are present. 
$b)$ Example of the wavefunction when spin and cavity dissipation are present  at the ``sweet spot'' in the phase diagram where $B_m = B_d$, and the system is initialized from an empty cavity.
$c)$ Example of the wavefunction when only spin dissipation is present. 
$d)$ Example of the wavefunction when only cavity dissipation is present.} \label{fig:QuantumSimulationsDissipation}
}
\end{figure}

To learn about the behaviour of the dissipative system, we simulated the evolution of the system using the above master equation. 
The Master equation is integrated stochastically, using the Stochastic Schroedinger Equation (SSE) method~\cite{Molmer1993}.
Doing this, we obtain trajectories of the system that can be seen as representative of the actual time evolution in an experiment.

To link our results to the non-dissipative case covered in Secs.~\ref{sec:NonDissNumerics}~and~\ref{sec:BoundaryBehaviour} (see Figs.~\ref{fig:PhaseDiagram}b and \ref{fig:QuantumSimulationsNoDissipation}abc), we use the same model parameters as in these sections. 
Specifically, we set $g = 2 \Omega/B_0$, $L=\frac{\hbar}{2}$,  $B_m=15B_0$, $B_d=8B_0$, where, as in Sec.~\ref{sec:FC:QuantumNoDissipation}, we use the field $B_0$ and the driving frequency $\Omega$ to set the physical scales for the simulation (see Sec.~\ref{sec:ModelIntroduction} for a physical definition of $B_0$).
We include the first $600$ photon states in the simulation, and set
  $\gamma_s=0.001/T$, $\gamma_c = 0.00053/T$, where $T \equiv 2\pi/\Omega$ denotes the period of the driving field.
The choice of $\gamma_c$ above implies that the cavity's photon emission rate (given by $2 \pi \gamma_c \langle n\rangle $, where $\langle n\rangle$ is the cavity photon number)  is exactly compensated by the topological pumping from the drive (given by $ L\Omega / \pi \hbar)$, when the cavity holds  $300$ photons. 

We initialized the cavity  in a coherent state with center at  $ n= 100$ photons and phase zero, and initially aligned  the spin along the $x$-axis, perpendicularly to the initial net field. 
In Fig.~\ref{fig:QuantumSimulationsDissipation}a, we show the evolution of a single, randomly picked realization of the SSE, obtained by solving the master equation in Eq.~\eqref{eq:RhoMasterEquation1} for the model, for the first $2000$ periods.
As can be seen, the system rapidly reaches a steady state where number of photons in the cavity mode fluctuates around $300$ (indicated by the dashed blue line), in agreement with our expectations.
The fluctuations of the system around this steady state can be seen as arising from quantum noise. 
The data in Fig.~\ref{fig:QuantumSimulationsDissipation}a  demonstrate that cavity and spin dissipation can stabilize a steady state in the system where the cavity emits photons at the topologically-quantized rate $1/T$. 
 The persistence of the topological pumping effect in the presence of dissipation thus  demonstrates that the effect does not rely on wave function coherence for its stability.
 
Note furthermore  that the spin dissipation leads to an alignment of the spin with the field; hence there is no reversed pumping in the system, unlike the non-dissipative case (see, e.g. Fig.~\ref{fig:QuantumSimulationsNoDissipation}a).
Even though the spin in Fig.~\ref{fig:QuantumSimulationDissipation}a was initially aligned perpendicularly to the initial net field, the photon number exclusively increases, unlike the non-dissipative case, where the same situation led to the generation of a ``cat'' state (see Fig.~\ref{fig:QuantumSimulationsNoDissipation}b). 
Thus, spin and cavity dissipation causes the highly-entangled ``cat'' discussed in Sec.~\ref{sec:FC:QuantumNoDissipation} to die. 
By forcing the spin to align itself  with the field, spin dissipation hence stabilizes the energy transfer effect compared to the non-dissipative case (see also discussion of Fig.~\ref{fig:QuantumSimulationsDissipation}c below).

Having established that the topological energy transfer is stabilized by dissipation, we  simulated the model at the special point in the phase diagram $B_m = B_d $, where the  system supported topological pumping from an  empty cavity in the non-dissipative case (see  Sec.~\ref{sec:EmptyCavity}).
Specifically, we included dissipation in the model studied in Sec.~\ref{sec:EmptyCavity}, i.e., with parameters $B_m=B_d = 10B_0$, $ g = 2\Omega/B_0$, $\omega = \Omega /\varphi$).
The dissipation strengths were set to  $\gamma_s = 0.001/T,\gamma_c = 0.00064/T$. 
The choice of $\gamma_c$ leads to a potential steady state with $\frac{1}{2\pi \gamma_c T}= 250$ photons. 
We initialized the system in an empty cavity state, and aligned the spin along the $x$-axis.
Fig.~\ref{fig:QuantumSimulationsDissipation}b shows the subsequent evolution for a single, randomly picked realization of the SSE. 
As can be seen, the photon number in the cavity increases to the steady-state value of $200$ photons, around which it subsequently fluctuates, similar to the behaviour in Fig.~\ref{fig:QuantumSimulationsDissipation}a. 
In contrast to Fig.~\ref{fig:QuantumSimulationsDissipation}a, however, the cavity was in this initialized in the vacuum.
This demonstrates that dissipation does not prevent the topological energy transfer to arise in an empty cavity.

To explore the effects of cavity and spin dissipation separately, we finally studied the system in the case where  only spin dissipation is present, and  in the case where only cavity dissipation is present. 
As in the beginning of this subsection, we again investigated the  model studied in Secs.~\ref{sec:NonDissNumerics}-\ref{sec:BoundaryBehaviour}, but with different choices for the dissipation strengths,  and with $700$ photon states included in the Hilbert space.  

To explore the effect of pure spin dissipation, we  set $\gamma_c= 0$, and $\gamma_s = 0.001/T$.
Fig.~\ref{fig:QuantumSimulationsDissipation}c shows the resulting evolution of the wave function for a single realization of the SSE. 
As can be seen, the photon number of the cavity increases at the linear rate $1/T$, until the wave function reaches the upper phase boundary at $n=529$ photons. 
After this point, the system's wave function  fluctuates around the upper phase boundary for the remainder of the simulation. 
Note that the wave function does not ``smear out'' over time, in contrast to the non-dissipative case (see Fig.~\ref{fig:QuantumSimulationsNoDissipation}c); 
the decoherence due to spin dissipation penalizes ``cat'' states that have a large spread in photon number and spin.
Moreover, the gradual Landau-Zener tunneling which occurs in the unitary case (see Fig.~\ref{fig:QuantumSimulationsNoDissipation}ac) is replaced with discrete quantum ``jumps'' near the upper phase boundary (horizontal dashed line), where the spin effectively flips instantaneously due to quantum noise. 
 At each jump, the direction of the photon transfer switches between $1/T$ and $-1/T$, and for most of the simulation, does not take value in between. 
Note that spin dissipation causes the system to remain confined to the upper boundary:  the spin can only Landau-Zener tunnel to anti-alignment with the field near the phase boundary, where the instantaneous energy gap  is comparable to the driving frequency (see discussion in Sec.~\ref{sec:FC:QuantumNoDissipation}). Away from this region,  the spin will  eventually align itself with the field, due to dissipation. As a result, the photon number will always increase below the phase boundary, and the system will thus remain confined near the upper boundary, as is the case in Fig.~\ref{fig:QuantumSimulationsDissipation}c.

We finally probe the dynamics of the system in the case where only cavity dissipation is present ($\gamma_s= 0 , \gamma_c = 0.00053 /T$).
The resulting data are shown in Fig.~\ref{fig:QuantumSimulationsDissipation}d for a single realization of the SSE. 
In this case, the system initially reaches the steady state of $300$ photons, where it remains confined for around $2000$ periods. 
However, after around $2100$ periods, quantum fluctuations cause the system to collapse at to a stable state outside the topological regime, where the cavity gradually depletes. 

\section{Classical nature of the effect}
\label{sec:Classical}
\begin{figure}
\includegraphics[width=\columnwidth]{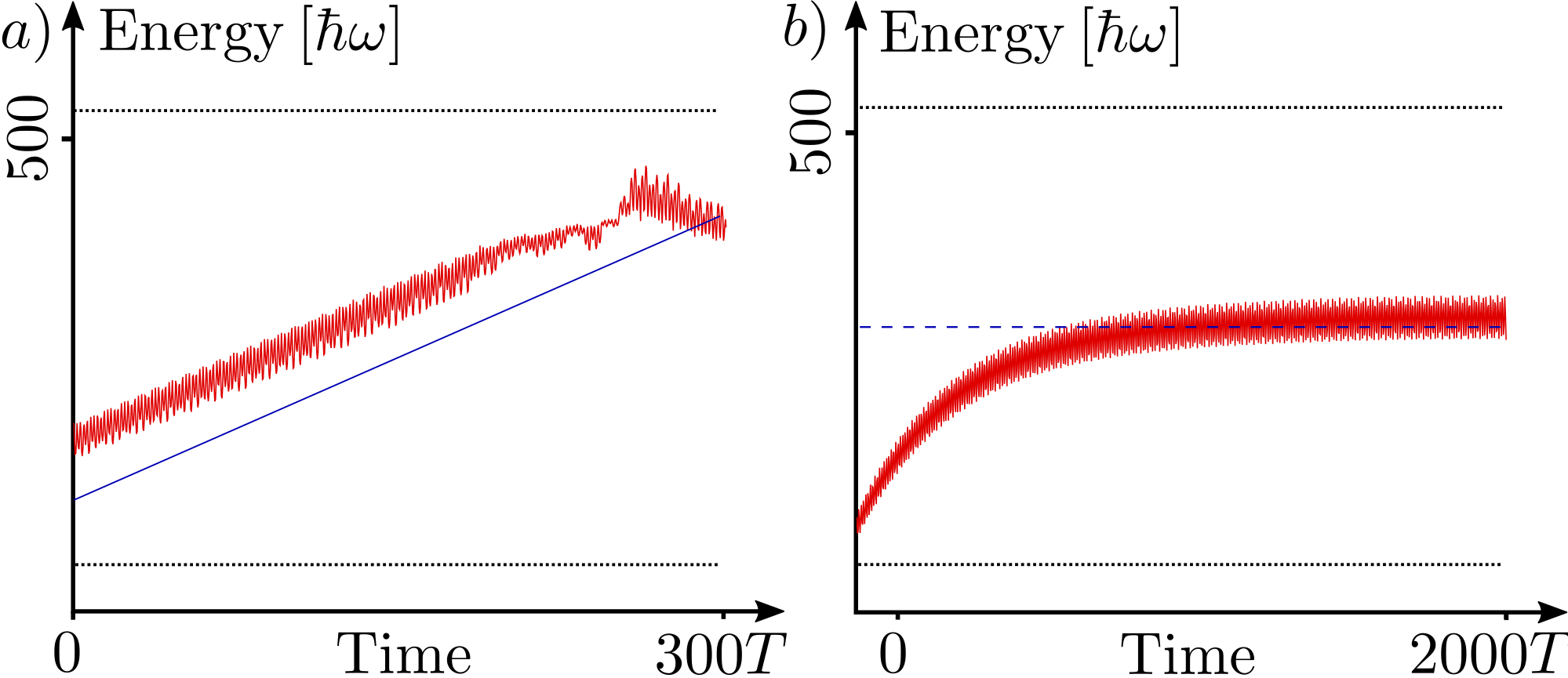}
\caption{
a) Evolution of the  cavity field energy for the classical model in Eq.~\eqref{eq:ClassicaHamiltonian}, with parameters and initialization corresponding to  the quantum system in Fig.~\ref{fig:PhaseDiagram}b -- see Sec.~\ref{sec:Classical} for details. 
Blue line  indicates  slope consistent with the topological  pumping rate $\dot E_c = L \omega \Omega /\pi$. 
Dashed lines indicate topological phase boundaries identified for the quantum case in Sec.~\ref{sec:FC:QuantumNoDissipation}.
b) Classical evolution of the  system in the presence of cavity and spin dissipation (see main text for details). 
Blue line indicates steady state at $E_c = 300 \hbar \omega$ expected from emission of energy at the topological pumping rate. }
\label{fig:Classical}
\end{figure}

The stability of the topological energy transfer in the presence of noise and dissipation, which was established above,  ints that the topological pumping effect has a classical counterpart. 
In this section we verify this intuition explicitly by a numerical simulation of the classical Hamiltonian  that corresponds to the model [Eq.~\eqref{eq:Hamiltonian0}].

With only a single mode present, the  classical Hamiltonian  in  Eq.~\eqref{eq:Hamiltonian0} depends on the  variables $B_c, \phi,$ and $ \vec L=(L_x,L_y,L_z)$, where $B_c $ and $\phi$ denote the amplitude and phase of the cavity mode. 
The classical amplitude and phase have the Poission bracket\cite{FN:PoissonBracket}
 $\{\phi,B_c\}=\frac{1}{2B_c}\frac{\omega\nu}{V}$,
 while the angular moment obeys the usual Poisson bracket relations $\{L_i, L_j\} = \epsilon_{ijk} L_k$.
As explained in Sec.~\ref{sec:ModelIntroduction}, the cavity field variables correspond  to the photonic creation and annihilation operators through the relation $B_0  a \sim B_c e^{-i\phi}$, where the  cavity-dependent magnetic field scale $B_0 \equiv \sqrt{{\omega \mu_0 \hbar }/V}$ was introduced  below Eq.~\eqref{eq:Hamiltonian0}. 
In terms of the classical variables above, the Hamiltonian reads
\be 
H(B_c, \phi, \vec L, t) = \frac{V }{\mu}B_c^2  + g\, \vec B(B_c,\phi,t) \cdot \vec L,
\label{eq:ClassicaHamiltonian}
\ee 
where  
\begin{eqnarray*}
 B_x(B_c,\phi,t) &=& B_c \sin(\phi) \\
 B_y (B_c, \phi,t) &=& B_d \sin(\Omega t ) \\
 B_z(B_c, \phi,t) &=& B_m-  B_d \cos (\Omega t)  - B_c \cos(\phi) .
\end{eqnarray*}
The energy of the cavity field is given by $E_c \equiv \frac{V}{\mu} B_c^2$, and  the dimensionless quantity $E_c /\hbar\omega$ corresponds to the number of photons in the cavity in the quantum mechanical case.


To explore the nature of the topological pumping effect, we numerically
solve the equations of motion for the classical Hamiltonian, which, for any of the variables $A$ above, can be expressed as $\dot A = \{A, H\}$. 
 We use the parameters that correspond to the quantum Hamiltonian studied in the previous sections (see, e.g., Sec.~\ref{sec:NonDissNumerics}).
Specifically, we set $g =2 \Omega/B_0$, $B_m = 15B_0$, $B_d = 8B_0$,  $\omega = \Omega /\varphi$, and chose the magnetic particle  to have angular moment $|\vec L |  =\hbar /2$. 
We initialize the cavity in the state $(B_c, \phi)=(14.1B_0,0)$, equivalent to  $200$ photons in the  quantum mechanical case, and initially aligned the angular moment along the $z$-axis (along the net initial  field), corresponding to the quantum initial state explored in Sec.~\ref{sec:NonDissNumerics} (see Fig.~\ref{fig:PhaseDiagram}b). 
In Fig.~\ref{fig:Classical}a, we  plot the resulting evolution of the cavity field energy $E_c $ (in units of $\hbar \omega$)  for the first $300$ driving periods. 
The  evolution of the classical field energy in Fig.~\ref{fig:Classical}a bears a clear resemblance to the evolution of the field energy in the quantum case (Fig.~\ref{fig:PhaseDiagram}b). 
In particular, the energy in the cavity increases at the constant quantized rate $\dot E_c= \omega \Omega L /2\pi$ (indicated by dashed blue line).
This behavior persists until the cavity energy reaches approximately $E_c = 530 \hbar \omega$,  at which the topological phase boundaries were identified in the quantum case (see discussion in Sec.~\ref{sec:NonDissNumerics}). 

The evolution in Fig.~\ref{fig:Classical}a shows that the topological energy transfer is  present in the fully classical model [Eq.~\eqref{eq:ClassicaHamiltonian}.
This conclusion can indeed be rigorously verified with a detailed analysis of the classical model~\cite{FollowUpWork}   (such an analysis is beyond the scope of this work, however).

We now verify the stability of the topological energy transfer  in the classical model  persists when dissipation is present. 
To demonstrate this, we introduced cavity and spin dissipation  in the equations of motion for $B_c$, $\phi$, and $\vec L$. 
Spin dissipation was modelled as a Gilbert damping  term: $\dot {\vec L} = \{\vec L, H \} - \frac{\gamma_s }{L} \vec L \times \dot{\vec L}$, where
 $\gamma_s$ is a dimensionless number which sets the spin dissipation strength. 
Cavity dissipation was included by modifying the equations of motion for $B_c$ such that $\dot  B_c  = \{B_c,H\} - \frac{\gamma_c}{2} B_c$, where $\gamma_c$ denotes the cavity dissipation strength. 
By explicit computation, one can verify that a nonzero value of $\gamma_c $ results in emission of energy from the  cavity at the rate  $\dot E_{\rm emission} = \gamma _cE_c$.

We solved the equations of motion with $\gamma_c=0.0033/T$, and $\gamma_s =0.001$, while parameters and  the initial state were chosen identical to the  realization studied above.
In Fig.~\ref{fig:Classical}b, we plot the resulting evolution of the cavity field energy. 
As can be seen, the system quickly reaches a steady state  with cavity energy $ E_c=300\hbar \omega$.
Following the above paragraph, this steady-state value results in emission of energy from the cavity at the rate $\dot E_{\rm emission} = \hbar \omega /T$. 
This value is   identical to the topological pumping rate $\Omega \omega  L /\pi$ derived in the previous sections. 
Hence, the ``lasing'' steady state, which was described in Sec.~\ref{sec:FC:QuantumWithDissipation}, is also a robust feature in the  classical limit of the model. 

\section{Multiple cavity modes}
\label{sec:MultipleModes}
\begin{figure}
\begin{center}
\includegraphics[width=1\columnwidth]{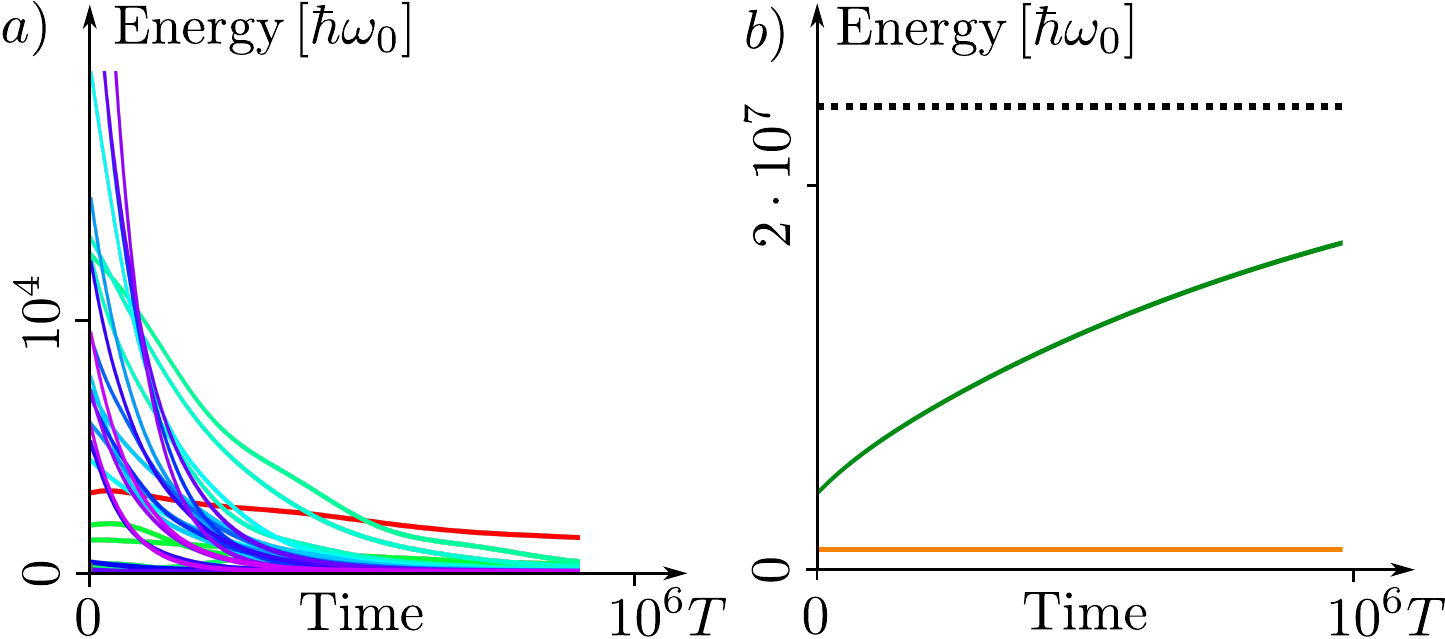}
\caption{Energy  stored in the cavity modes as a function of time, in a setting where the first harmonic is in the topological regime, while  the remaining modes are initially only weakly excited. 
See main text for details. 
a) Energies of modes $\pm  2 \ldots \pm 15$ (green-purple curves) and mode $-1$ (red curve). 
b) Energy of  mode $1$ (green curve). }
\label{fig:MultipleModes}
\end{center}
\end{figure}

Up to this point, we have for simplicity assumed that the magnetic particle is only coupled to a single (circularly-polarized) cavity mode. 
However, a realistic cavity will support a large number of distinct modes, which, in general, will all couple to the magnetic particle with nonzero strength.
Each additional mode adds a extra synthetic dimension, which, in particular, is expected to change its topological classification and features it can support.
In this section we demonstrate that  topological frequency conversion  nevertheless can persist even  when these additional cavity modes are taken into account.

We consider here the case where the cavity is effectively one-dimensional. 
In this case, for each integer $m$,
the cavity supports  two  distinct modes with frequency $\omega_n \equiv n\omega_0$, where $\omega_0$ denotes the fundamental frequency of the cavity.
The two modes have clockwise and counterclockwise polarization, respectively. 
 We label the clockwise-polarized mode  $n$, and the counterclockwise-polarized mode $-n$.

The magnetic field from mode $n$ is given by $\vec B_n = B_n(0,\sin\phi_n,-\cos\phi_n)$, where $\phi_n$ and $B_n$ respectively denote the phase and amplitude of mode $n$.
In terms of the fields $\{\vec B_n\}$, the Hamiltonian of the system can be written 
\be 
H(t)=\sum_{n } \left(\frac{V}{\mu_0}  \vec B_n^2+  g_n \vec B_n \cdot \vec L\right)+   g_d (\vec B_d (t)+ \vec B_m) \cdot \vec L,
\label{eq:MMHamiltonian}
\ee 
where $\vec B_d(t) = (B_d \sin\Omega t,0, -B_d \cos \Omega t )$ denotes the driving field, and $\vec B_m = (0,0,B_m)$ denotes the static Zeeman field.
The coupling $g_n$ may depend on the frequency of the mode. 

We expect that significant excitation of multiple cavity modes will lead to chaotic behaviour of the spin and the system, and to a breakdown of the topological energy pumping. 
We are therefore interested in finding a regime, where only a single mode $n_0$ is significantly excited. 
In this case, the discussion of the previous sections applies, and the system may undergo topological frequency conversion.   

To see how such a suppression of spurious modes can be achieved, note that mode $n$ is excited by the motion of the moment $\vec L$  through the coupling $g_n B_n \cdot L$, while it decays due to dissipation, which is controlled by the (mode-dependent) dissipation strength $\gamma_n$.
The competition of the excitation and dissipation  determines the mean amplitude of the mode.
The spurious modes can thus be prevented from obstructing the pumping  by maximizing their dissipation rates $\{\gamma_n\}$ and minimizing their couplings $\{g_n\}$. 

The dissipation rate $\gamma_n$ can be made strongly frequency-dependent  by letting the cavity's mirror reflection coefficient depend on frequency. 
Such a frequency dependence can for instance be realized with Fabry-P\'erot devices, which may effectively suppress all undesirable harmonics in the cavity~\cite{Perot1899,Hernandez1986,Ismail2016FabryPerot}. 
The coupling $g_n$  to high modes can be suppressed by distributing the magnetic particles  over an extended spatial region in the cavity.
 In this case, the effective coupling $g_n$ to modes with higher frequency (corresponding to shorter wavelength)    decreases as $|n|^{-1}$.

Finally, since the spurious modes are excited from the motion of the angular moment $\vec L$, the amplitudes of a spurious  mode are highly dependent on whether or not the oscillations of $\vec L$ are resonant with the frequency of the modes. 
Since the motion of $\vec L$ should mainly be driven by the oscillation of the driving mode and the pumped mode $n_0$, its frequency spectrum in the pumping regime should  feature sharp peaks at integer-multiple combinations of $\omega_{n_0}$ and $\Omega$. 
Thus the amplitudes of the spurious modes can be dramatically decreased if their frequencies $\{\omega_n\}$  are even slightly detuned from those peaks, e.g., if $\omega_n = n \omega_0 + \delta \omega$ for some nonzero (but small) $\delta \omega$.
Such detuning can be realized by inserting phase-shifting mirror in the cavity~\cite{Born1999,Aurand2011}.

\subsection{Numerical study}
\label{sec:MM:Numerics}
We now demonstrate, through  numerical simulations, that it is possible to realize the topological pumping effect in a multi-mode cavity using the strategies discussed above. 
Specifically, we show that a regime exists where mode $1$ undergoes topological energy pumping, while all other cavity modes effectively  remain unexcited. 

Since Sec.~\ref{sec:Classical} demonstrated that the energy pumping effect is of classical nature, we may simulate the system accurately using its classical equations of motion. 
We include the first 15 harmonics in the model, along with their polarization-reversed partners.

To avoid resonances with the spurious modes, we detuned the  frequencies of the modes  slightly from the integer multiples of the fundamental frequency.
Specifically, we set $\omega_n = n \omega_0 + \delta \omega$, 
with $\delta \omega= 0.05 \omega_0$, where the ``bare'' fundamental frequency $\omega_0$ is given below. 
Additionally, we let the coupling strengths $\{g_n\}$  depend on the mode index  as $|n|^{-1}$.
Finally, we let the dissipation rate for a mode depend linearly on the mode's frequency such that  $\gamma_n  = |n|\gamma_1$. 

Using the driving frequency $\Omega$ and the field strength $B_0$ to set the  scales for our simulation, we set $L/L_0 = 3\hbar $,  $\omega_0 = {\Omega}/{\varphi} $ where $\varphi= (\sqrt 5-1 )/2$.
The dimensionless dissipation rate for the spin is set to $\gamma_s = 0.01$, while   $\gamma_1$ was set to $ 1.59 \cdot 10^{- 7} \Omega$, which leads to an expected steady state at energy $E_0  = \frac{L \omega_1 \Omega}{  \pi  \gamma_1} = 6 \cdot 10^6\hbar \omega$. 
We set $g_d= g_1 = 0.0047 \Omega/ B_0$.
Finally, we set
$B_m=2,\!000 B_0$, $A_d = 1,\! 400 B_0 $.
From an analysis similar to that made in Sec.~\ref{sec:NonDissNumerics}, we expect that mode $1$ is in the topological pumping regime if its  amplitude lies in the interval between $ B_- $ and $B_+$, where $B_- \equiv|B_ m-B_d|=600 B_0$, and $B_+  \equiv |B_m+B_d| = 3400 B_0$.

To investigate whether the system supports topological frequency conversion, we initialized  mode $1$ in the state $B_1 = 1000 B_0$, $\phi_0 = 0$, while the initial amplitudes of the remaining 29 modes  are  randomly distributed in the interval between $0$ and $32B_0$. 
The phases of all modes are initially set to zero, and the angular moment $\vec L$ was initially aligned along the $z$-axis. 
Starting from this initial state, we numerically solve the  classical equations of motion, and plot in Fig.~\ref{fig:MultipleModes}a,b  the resulting evolution  of the modes' energies over  the first $10^6$ driving periods. 
The red curve in panel $a)$ indicates the energy of mode $-1$, while the dark green curve on panel $b)$ indicates the energy  of mode $1$. 
The remaining curves in panel $a)$ indicate the energies of the other modes, with 
green for lower harmonics, and blue or purple for higher harmonics.

As can be seen in panel $a)$, the energies of  all   modes other than mode $1$ decay to near zero\cite{FN:TimeReversedPartner},
 over a few hundred thousand driving periods, with the higher-frequency  modes decaying more rapidly.  
If additional modes were included in the simulation, we expect their amplitudes to decay even faster.
The small amplitudes of the spurious modes allow mode $1 $ to undergo topological pumping: 
as  panel b) shows, the energy of mode $1$  approaches the steady state value of $E_0 = 6 \cdot 10^6 \hbar \omega_0$ (dashed line) (the energy of mode $1$ might eventually settle at a value slightly below   this value, due to the energy loss to the other modes). 
As discussed above, this value is  consistent with  the cavity being topologically pumped at the universal rate $\omega_0\Omega L/\pi$ .

\section{Discussion}
\label{sec:FC:Discussion}
In this work, we demonstrated the  robustness of the topological energy transfer between two circularly polarized modes that was proposed in Ref.~\onlinecite{Martin2017}. 
We studied the setup under the experimentally relevant settings where one of the modes is externally driven, while the other is a dynamical cavity mode. 
Using a new   master equation technique, we established that the effect is stable in the presence of noise and dissipation.
In particular, cavity dissipation, in the form of a semitransparent mirror, can stabilize a steady state where cavity photons are emitted at the quantized rate of one per driving period. 
The dissipation due to the magnetic particle's motion  may even  add to the robustness of the effect, by keeping the magnetic moment aligned with the instantaneous field. 
Finally, we find that the effect can be realized even if the cavity is initially empty.

The robustness of  topological energy transfer  is further reflected in the fact that the effect can be understood  as a {\it classical} phenomenon. 
Hence, it does not rely on coherence of wave functions for its stability, and can be realized in noisy, macroscopic systems. 
Remarkably, our results thus reveal that simple classical systems can  exhibit highly nontrivial topological effects. 

The fact that the topological energy transfer  is a  classical phenomenon  means that the effect can be effectively simulated by classical equations of motion. 
This allows for simulation of the effect under much more complex settings than if it were a purely quantum mechanical effect.
We used this observation to to show that the effect can persist even if multiple modes are present in the cavity. 
This opens up the possibility for realistic experimental realizations. 


\paragraph*{Experimental realization and future work --- }
To implement the topological energy transfer experimentally, one can exploit the fact that the effect has a classical incarnation. 
Hence it can be realized in macroscopic, mechanical systems -- for instance, using a fast-spinning gyroscope (instead of spin), coupled to mechanical harmonic oscillators (rather than electromagnetic modes), such as pendulums or springs.

Alternatively, the model  can be realized with an actual magnetic moment coupled to electromagnetic modes.
Importantly, the  magnetic particle's isotropic gyromagnetic ratio $g$ in this case sets the range of  frequencies where topological frequency conversion can realistically arise:
as the discussion in Sec.~\ref{sec:TopologicalFrequencyConversion} shows,  
topological frequency conversion requires the  magnetic particle's   precession frequency $g|\vec B| $ to be  large compared to  the driving and cavity modes' frequencies $\omega$ and $\Omega$  (here $|\vec B|$ denotes the magnitude of the instantaneous magnetic field).  
Using that the mode's field intensity is given by $I= \vec B^2 c /\mu_0$ (where $c$ is the speed of light), we conclude that  
topological frequency conversion  requires the field intensity   to  be of order  $I_{\rm min} = \frac{\omega_{0}^2 c }{ g^2 \mu_0}$  or larger,  where $\omega_{0}$ denotes the largest of the frequencies $\omega$ and $\Omega$.

The largest    isotropic gyromagnetic ratios that are currently known are comparable to the gyromagnetic ratio of the electron $  g_{e} \approx 1.76 \cdot 10^{11} {\rm Hz}\, {\rm T}^{-1}$ (i.e., corresponding to $g$-factors of $2$).
 As discussed in Ref.~\onlinecite{Martin2017}, such  values of $g$ can for example be achieved with  an Yttrium-Iron garnet (YIG) sphere
 ~\cite{Tyryshkin2003}.
Using  $g=g_e$, the minimal intensity required for topological frequency conversion (as a function of frequency) can be expressed as $I _{\rm min} [{\rm W}/{\rm cm}^2] \approx (\omega_{0}[ {\rm MHz}])^2$. 
In this way, rather strong  radiation intensities of order $100\, {\rm W }/{\rm cm}^2$  allows for topological frequency conversion up to the $10 \, {\rm MHz}$ range, while topological  conversion of ${\rm THz}$ frequencies requires radiation intensities of $10^{12} {\rm W}/{\rm cm}^2$.

The above estimates shows that achieving frequency conversion at high frequencies may be difficult using magnetic couplings. An alternative is to couple to the electric field component of electromagnetic waves, which due to a generically stronger coupling may allow to go beyond  beyond the ${\rm GHz}$ range.  
A detailed discussion of such an implementation is left for future studies. 
One possibility is to replace the magnetic particle discussed above with the orbital (pseudospin) degree of freedom in a Weyl semimetal~\cite{Singh2012,Wan2011,Burkow2011a,Burkov2011b}, as was also mentioned in Ref.~\onlinecite{Martin2017}.
Note that the quadratic dependence of the energy transfer rate on the frequencies ($\dot E = \omega_1 \omega_2 L/\pi$) mean that even a tiny net moment can result in macroscopic energy transfer rates, for sufficiently large frequencies.

The discussion in Sec.~\ref{sec:MultipleModes} shows that the effectiveness of the topological frequency conversion is highly dependent on the properties of the electromagnetic cavity.
In particular, the effect depends strongly on the  spectrum of modes in the cavity, and their dissipation rates. 
A detailed discussion of how a cavity with suitable properties can be implemented is beyond the scope of this work, and is left for the future. 

{\it Aknowledgments ---}
FN aknowledges support from the Villum foundation and the Danish National Research Foundation. Work at Argonne National Laboratory was supported by the Department of Energy, Office of Science,  Materials Science and Engineering Division.
GR is grateful for support from the Institute of Quantum Information and Matter, an NSF Frontier center funded by the Gordon and Betty Moore Foundation, the Packard Foundation, and from the ARO MURI W911NF-16-1-0361 “Quantum Materials by Design with Electromagnetic Excitation'' sponsored by the U.S. Army. IM and GR are grateful for the hospitality of the Aspen Center for Physics, supported by National Science Foundation grant PHY-1607611. 
\bibliographystyle{apsrev}
\bibliography{Bibliography.bib}

\begin{thebibliography}{62}
\expandafter\ifx\csname natexlab\endcsname\relax\def\natexlab#1{#1}\fi
\expandafter\ifx\csname bibnamefont\endcsname\relax
  \def\bibnamefont#1{#1}\fi
\expandafter\ifx\csname bibfnamefont\endcsname\relax
  \def\bibfnamefont#1{#1}\fi
\expandafter\ifx\csname citenamefont\endcsname\relax
  \def\citenamefont#1{#1}\fi
\expandafter\ifx\csname url\endcsname\relax
  \def\url#1{\texttt{#1}}\fi
\expandafter\ifx\csname urlprefix\endcsname\relax\def\urlprefix{URL }\fi
\providecommand{\bibinfo}[2]{#2}
\providecommand{\eprint}[2][]{\url{#2}}

\bibitem[{\citenamefont{Thouless et~al.}(1982)\citenamefont{Thouless, Kohmoto,
  Nightingale, and den Nijs}}]{Thouless1982TKNN}
\bibinfo{author}{\bibfnamefont{D.~J.} \bibnamefont{Thouless}},
  \bibinfo{author}{\bibfnamefont{M.}~\bibnamefont{Kohmoto}},
  \bibinfo{author}{\bibfnamefont{M.~P.} \bibnamefont{Nightingale}},
  \bibnamefont{and} \bibinfo{author}{\bibfnamefont{M.}~\bibnamefont{den Nijs}},
  \bibinfo{journal}{Phys. Rev. Lett.} \textbf{\bibinfo{volume}{49}},
  \bibinfo{pages}{405} (\bibinfo{year}{1982}),
  \urlprefix\url{http://link.aps.org/doi/10.1103/PhysRevLett.49.405}.

\bibitem[{\citenamefont{Bernevig et~al.}(2006)\citenamefont{Bernevig, Hughes,
  and Zhang}}]{Bernevig2006}
\bibinfo{author}{\bibfnamefont{B.~A.} \bibnamefont{Bernevig}},
  \bibinfo{author}{\bibfnamefont{T.~L.} \bibnamefont{Hughes}},
  \bibnamefont{and} \bibinfo{author}{\bibfnamefont{S.-C.} \bibnamefont{Zhang}},
  \bibinfo{journal}{Science} \textbf{\bibinfo{volume}{314}},
  \bibinfo{pages}{1757} (\bibinfo{year}{2006}).

\bibitem[{\citenamefont{Kane and Mele}(2005)}]{Kane2005_1}
\bibinfo{author}{\bibfnamefont{C.~L.} \bibnamefont{Kane}} \bibnamefont{and}
  \bibinfo{author}{\bibfnamefont{E.~J.} \bibnamefont{Mele}},
  \bibinfo{journal}{Phys. Rev. Lett.} \textbf{\bibinfo{volume}{95}},
  \bibinfo{pages}{146802} (\bibinfo{year}{2005}).

\bibitem[{\citenamefont{Fu and Kane}(2006)}]{Fu2006}
\bibinfo{author}{\bibfnamefont{L.}~\bibnamefont{Fu}} \bibnamefont{and}
  \bibinfo{author}{\bibfnamefont{C.~L.} \bibnamefont{Kane}},
  \bibinfo{journal}{Phys. Rev. B} \textbf{\bibinfo{volume}{74}},
  \bibinfo{pages}{195312} (\bibinfo{year}{2006}).

\bibitem[{\citenamefont{Kitaev}(2009)}]{Kitaev2009}
\bibinfo{author}{\bibfnamefont{A.}~\bibnamefont{Kitaev}}, \bibinfo{journal}{AIP
  Conference Proceedings} \textbf{\bibinfo{volume}{1134}}, \bibinfo{pages}{22}
  (\bibinfo{year}{2009}).

\bibitem[{\citenamefont{Ryu et~al.}(2010)\citenamefont{Ryu, Schnyder, Furusaki,
  and Ludwig}}]{Schnyder2010}
\bibinfo{author}{\bibfnamefont{S.}~\bibnamefont{Ryu}},
  \bibinfo{author}{\bibfnamefont{A.~P.} \bibnamefont{Schnyder}},
  \bibinfo{author}{\bibfnamefont{A.}~\bibnamefont{Furusaki}}, \bibnamefont{and}
  \bibinfo{author}{\bibfnamefont{A.~W.~W.} \bibnamefont{Ludwig}},
  \bibinfo{journal}{New Journal of Physics} \textbf{\bibinfo{volume}{12}},
  \bibinfo{pages}{065010} (\bibinfo{year}{2010}).

\bibitem[{\citenamefont{K\"onig et~al.}(2007)\citenamefont{K\"onig, Wiedmann,
  Br\"une, Roth, Buhmann, Molenkamp, Qi, and Zhang}}]{Konig2007_1}
\bibinfo{author}{\bibfnamefont{M.}~\bibnamefont{K\"onig}},
  \bibinfo{author}{\bibfnamefont{S.}~\bibnamefont{Wiedmann}},
  \bibinfo{author}{\bibfnamefont{C.}~\bibnamefont{Br\"une}},
  \bibinfo{author}{\bibfnamefont{A.}~\bibnamefont{Roth}},
  \bibinfo{author}{\bibfnamefont{H.}~\bibnamefont{Buhmann}},
  \bibinfo{author}{\bibfnamefont{L.~W.} \bibnamefont{Molenkamp}},
  \bibinfo{author}{\bibfnamefont{X.-L.} \bibnamefont{Qi}}, \bibnamefont{and}
  \bibinfo{author}{\bibfnamefont{S.-C.} \bibnamefont{Zhang}},
  \bibinfo{journal}{Science} \textbf{\bibinfo{volume}{318}},
  \bibinfo{pages}{766} (\bibinfo{year}{2007}).

\bibitem[{\citenamefont{Hsieh et~al.}(2008)\citenamefont{Hsieh, Qian, Wray,
  Xia, Hor, Cava, and Hasan}}]{Hsieh2008_1}
\bibinfo{author}{\bibfnamefont{D.}~\bibnamefont{Hsieh}},
  \bibinfo{author}{\bibfnamefont{D.}~\bibnamefont{Qian}},
  \bibinfo{author}{\bibfnamefont{L.}~\bibnamefont{Wray}},
  \bibinfo{author}{\bibfnamefont{Y.}~\bibnamefont{Xia}},
  \bibinfo{author}{\bibfnamefont{Y.~S.} \bibnamefont{Hor}},
  \bibinfo{author}{\bibfnamefont{R.~J.} \bibnamefont{Cava}}, \bibnamefont{and}
  \bibinfo{author}{\bibfnamefont{M.~Z.} \bibnamefont{Hasan}},
  \bibinfo{journal}{Nature} \textbf{\bibinfo{volume}{452}},
  \bibinfo{pages}{970} (\bibinfo{year}{2008}), ISSN \bibinfo{issn}{0028-0836}.

\bibitem[{\citenamefont{Yao et~al.}(2007)\citenamefont{Yao, MacDonald, and
  Niu}}]{Yao2007}
\bibinfo{author}{\bibfnamefont{W.}~\bibnamefont{Yao}},
  \bibinfo{author}{\bibfnamefont{A.~H.} \bibnamefont{MacDonald}},
  \bibnamefont{and} \bibinfo{author}{\bibfnamefont{Q.}~\bibnamefont{Niu}},
  \bibinfo{journal}{Physical Review Letters} \textbf{\bibinfo{volume}{99}},
  \bibinfo{pages}{047401} (\bibinfo{year}{2007}), ISSN
  \bibinfo{issn}{0031-9007},
  \urlprefix\url{https://link.aps.org/doi/10.1103/PhysRevLett.99.047401}.

\bibitem[{\citenamefont{Oka and Aoki}(2009)}]{Oka2009}
\bibinfo{author}{\bibfnamefont{T.}~\bibnamefont{Oka}} \bibnamefont{and}
  \bibinfo{author}{\bibfnamefont{H.}~\bibnamefont{Aoki}},
  \bibinfo{journal}{Phys. Rev. B} \textbf{\bibinfo{volume}{79}},
  \bibinfo{pages}{081406} (\bibinfo{year}{2009}).

\bibitem[{\citenamefont{Inoue and Tanaka}(2010)}]{Inoue2010}
\bibinfo{author}{\bibfnamefont{J.-i.} \bibnamefont{Inoue}} \bibnamefont{and}
  \bibinfo{author}{\bibfnamefont{A.}~\bibnamefont{Tanaka}},
  \bibinfo{journal}{Phys. Rev. Lett.} \textbf{\bibinfo{volume}{105}},
  \bibinfo{pages}{017401} (\bibinfo{year}{2010}).

\bibitem[{\citenamefont{Kitagawa et~al.}(2010)\citenamefont{Kitagawa, Berg,
  Rudner, and Demler}}]{Kitagawa2010}
\bibinfo{author}{\bibfnamefont{T.}~\bibnamefont{Kitagawa}},
  \bibinfo{author}{\bibfnamefont{E.}~\bibnamefont{Berg}},
  \bibinfo{author}{\bibfnamefont{M.}~\bibnamefont{Rudner}}, \bibnamefont{and}
  \bibinfo{author}{\bibfnamefont{E.}~\bibnamefont{Demler}},
  \bibinfo{journal}{Phys. Rev. B} \textbf{\bibinfo{volume}{82}},
  \bibinfo{pages}{235114} (\bibinfo{year}{2010}).

\bibitem[{\citenamefont{Lindner et~al.}(2011)\citenamefont{Lindner, Refael, and
  Galitski}}]{Lindner2011}
\bibinfo{author}{\bibfnamefont{N.~H.} \bibnamefont{Lindner}},
  \bibinfo{author}{\bibfnamefont{G.}~\bibnamefont{Refael}}, \bibnamefont{and}
  \bibinfo{author}{\bibfnamefont{V.}~\bibnamefont{Galitski}},
  \bibinfo{journal}{Nat. Phys.} \textbf{\bibinfo{volume}{7}},
  \bibinfo{pages}{490} (\bibinfo{year}{2011}).

\bibitem[{\citenamefont{Jotzu et~al.}(2014)\citenamefont{Jotzu, Messer,
  Desbuquois, Lebrat, Uehlinger, Greif, and Esslinger}}]{Jotzu2014}
\bibinfo{author}{\bibfnamefont{G.}~\bibnamefont{Jotzu}},
  \bibinfo{author}{\bibfnamefont{M.}~\bibnamefont{Messer}},
  \bibinfo{author}{\bibfnamefont{R.}~\bibnamefont{Desbuquois}},
  \bibinfo{author}{\bibfnamefont{M.}~\bibnamefont{Lebrat}},
  \bibinfo{author}{\bibfnamefont{T.}~\bibnamefont{Uehlinger}},
  \bibinfo{author}{\bibfnamefont{D.}~\bibnamefont{Greif}}, \bibnamefont{and}
  \bibinfo{author}{\bibfnamefont{T.}~\bibnamefont{Esslinger}},
  \bibinfo{journal}{Nature} \textbf{\bibinfo{volume}{515}},
  \bibinfo{pages}{237} (\bibinfo{year}{2014}).

\bibitem[{\citenamefont{Jiang et~al.}(2011)\citenamefont{Jiang, Kitagawa,
  Alicea, Akhmerov, Pekker, Refael, Cirac, Demler, Lukin, and
  Zoller}}]{Jiang2011}
\bibinfo{author}{\bibfnamefont{L.}~\bibnamefont{Jiang}},
  \bibinfo{author}{\bibfnamefont{T.}~\bibnamefont{Kitagawa}},
  \bibinfo{author}{\bibfnamefont{J.}~\bibnamefont{Alicea}},
  \bibinfo{author}{\bibfnamefont{A.~R.} \bibnamefont{Akhmerov}},
  \bibinfo{author}{\bibfnamefont{D.}~\bibnamefont{Pekker}},
  \bibinfo{author}{\bibfnamefont{G.}~\bibnamefont{Refael}},
  \bibinfo{author}{\bibfnamefont{J.~I.} \bibnamefont{Cirac}},
  \bibinfo{author}{\bibfnamefont{E.}~\bibnamefont{Demler}},
  \bibinfo{author}{\bibfnamefont{M.~D.} \bibnamefont{Lukin}}, \bibnamefont{and}
  \bibinfo{author}{\bibfnamefont{P.}~\bibnamefont{Zoller}},
  \bibinfo{journal}{Phys. Rev. Lett.} \textbf{\bibinfo{volume}{106}},
  \bibinfo{pages}{220402} (\bibinfo{year}{2011}).

\bibitem[{\citenamefont{Rudner et~al.}(2013)\citenamefont{Rudner, Lindner,
  Berg, and Levin}}]{Rudner2013}
\bibinfo{author}{\bibfnamefont{M.~S.} \bibnamefont{Rudner}},
  \bibinfo{author}{\bibfnamefont{N.~H.} \bibnamefont{Lindner}},
  \bibinfo{author}{\bibfnamefont{E.}~\bibnamefont{Berg}}, \bibnamefont{and}
  \bibinfo{author}{\bibfnamefont{M.}~\bibnamefont{Levin}},
  \bibinfo{journal}{Phys. Rev. X} \textbf{\bibinfo{volume}{3}},
  \bibinfo{pages}{031005} (\bibinfo{year}{2013}).

\bibitem[{\citenamefont{Nathan and Rudner}(2015)}]{Nathan2015}
\bibinfo{author}{\bibfnamefont{F.}~\bibnamefont{Nathan}} \bibnamefont{and}
  \bibinfo{author}{\bibfnamefont{M.~S.} \bibnamefont{Rudner}},
  \bibinfo{journal}{New Journal of Physics} \textbf{\bibinfo{volume}{17}},
  \bibinfo{pages}{125014} (\bibinfo{year}{2015}), ISSN
  \bibinfo{issn}{1367-2630}.

\bibitem[{\citenamefont{Roy and Harper}(2016)}]{Roy2016}
\bibinfo{author}{\bibfnamefont{R.}~\bibnamefont{Roy}} \bibnamefont{and}
  \bibinfo{author}{\bibfnamefont{F.}~\bibnamefont{Harper}},
  \bibinfo{journal}{Physical Review B} \textbf{\bibinfo{volume}{94}},
  \bibinfo{pages}{125105} (\bibinfo{year}{2016}), ISSN
  \bibinfo{issn}{2469-9950},
  \urlprefix\url{https://link.aps.org/doi/10.1103/PhysRevB.94.125105}.

\bibitem[{\citenamefont{von Keyserlingk and
  Sondhi}(2016)}]{vonKeyserlingk2016a}
\bibinfo{author}{\bibfnamefont{C.~W.} \bibnamefont{von Keyserlingk}}
  \bibnamefont{and} \bibinfo{author}{\bibfnamefont{S.~L.}
  \bibnamefont{Sondhi}}, \bibinfo{journal}{Phys. Rev. B}
  \textbf{\bibinfo{volume}{93}}, \bibinfo{pages}{245145}
  (\bibinfo{year}{2016}).

\bibitem[{\citenamefont{Potter et~al.}(2016)\citenamefont{Potter, Morimoto, and
  Vishwanath}}]{Potter2016}
\bibinfo{author}{\bibfnamefont{A.~C.} \bibnamefont{Potter}},
  \bibinfo{author}{\bibfnamefont{T.}~\bibnamefont{Morimoto}}, \bibnamefont{and}
  \bibinfo{author}{\bibfnamefont{A.}~\bibnamefont{Vishwanath}},
  \bibinfo{journal}{Phys. Rev. X} \textbf{\bibinfo{volume}{6}},
  \bibinfo{pages}{041001} (\bibinfo{year}{2016}).

\bibitem[{\citenamefont{Else et~al.}(2016)\citenamefont{Else, Bauer, and
  Nayak}}]{Else2016b}
\bibinfo{author}{\bibfnamefont{D.~V.} \bibnamefont{Else}},
  \bibinfo{author}{\bibfnamefont{B.}~\bibnamefont{Bauer}}, \bibnamefont{and}
  \bibinfo{author}{\bibfnamefont{C.}~\bibnamefont{Nayak}},
  \bibinfo{journal}{Phys. Rev. Lett.} \textbf{\bibinfo{volume}{117}},
  \bibinfo{pages}{090402} (\bibinfo{year}{2016}),
  \urlprefix\url{http://link.aps.org/doi/10.1103/PhysRevLett.117.090402}.

\bibitem[{\citenamefont{Khemani et~al.}(2016)\citenamefont{Khemani, Lazarides,
  Moessner, and Sondhi}}]{Khemani16}
\bibinfo{author}{\bibfnamefont{V.}~\bibnamefont{Khemani}},
  \bibinfo{author}{\bibfnamefont{A.}~\bibnamefont{Lazarides}},
  \bibinfo{author}{\bibfnamefont{R.}~\bibnamefont{Moessner}}, \bibnamefont{and}
  \bibinfo{author}{\bibfnamefont{S.~L.} \bibnamefont{Sondhi}},
  \bibinfo{journal}{Phys. Rev. Lett.} \textbf{\bibinfo{volume}{116}},
  \bibinfo{pages}{250401} (\bibinfo{year}{2016}),
  \urlprefix\url{http://link.aps.org/doi/10.1103/PhysRevLett.116.250401}.

\bibitem[{\citenamefont{{Choi} et~al.}(2017)\citenamefont{{Choi}, {Choi},
  {Landig}, {Kucsko}, {Zhou}, {Isoya}, {Jelezko}, {Onoda}, {Sumiya}, {Khemani}
  et~al.}}]{Choi16DTC}
\bibinfo{author}{\bibfnamefont{S.}~\bibnamefont{{Choi}}},
  \bibinfo{author}{\bibfnamefont{J.}~\bibnamefont{{Choi}}},
  \bibinfo{author}{\bibfnamefont{R.}~\bibnamefont{{Landig}}},
  \bibinfo{author}{\bibfnamefont{G.}~\bibnamefont{{Kucsko}}},
  \bibinfo{author}{\bibfnamefont{H.}~\bibnamefont{{Zhou}}},
  \bibinfo{author}{\bibfnamefont{J.}~\bibnamefont{{Isoya}}},
  \bibinfo{author}{\bibfnamefont{F.}~\bibnamefont{{Jelezko}}},
  \bibinfo{author}{\bibfnamefont{S.}~\bibnamefont{{Onoda}}},
  \bibinfo{author}{\bibfnamefont{H.}~\bibnamefont{{Sumiya}}},
  \bibinfo{author}{\bibfnamefont{V.}~\bibnamefont{{Khemani}}},
  \bibnamefont{et~al.}, \bibinfo{journal}{Nature}
  \textbf{\bibinfo{volume}{543}}, \bibinfo{pages}{221} (\bibinfo{year}{2017}).

\bibitem[{\citenamefont{Zhang et~al.}(2017)\citenamefont{Zhang, Hess,
  Kyprianidis, Becker, Lee, Smith, Pagano, Potirniche, Potter, Vishwanath
  et~al.}}]{MonroeDTC}
\bibinfo{author}{\bibfnamefont{J.}~\bibnamefont{Zhang}},
  \bibinfo{author}{\bibfnamefont{P.~W.} \bibnamefont{Hess}},
  \bibinfo{author}{\bibfnamefont{A.}~\bibnamefont{Kyprianidis}},
  \bibinfo{author}{\bibfnamefont{P.}~\bibnamefont{Becker}},
  \bibinfo{author}{\bibfnamefont{A.}~\bibnamefont{Lee}},
  \bibinfo{author}{\bibfnamefont{J.}~\bibnamefont{Smith}},
  \bibinfo{author}{\bibfnamefont{G.}~\bibnamefont{Pagano}},
  \bibinfo{author}{\bibfnamefont{I.-D.} \bibnamefont{Potirniche}},
  \bibinfo{author}{\bibfnamefont{A.~C.} \bibnamefont{Potter}},
  \bibinfo{author}{\bibfnamefont{A.}~\bibnamefont{Vishwanath}},
  \bibnamefont{et~al.}, \bibinfo{journal}{Nature}
  \textbf{\bibinfo{volume}{543}}, \bibinfo{pages}{217} (\bibinfo{year}{2017}).

\bibitem[{\citenamefont{Titum et~al.}(2016)\citenamefont{Titum, Berg, Rudner,
  Refael, and Lindner}}]{AFAI}
\bibinfo{author}{\bibfnamefont{P.}~\bibnamefont{Titum}},
  \bibinfo{author}{\bibfnamefont{E.}~\bibnamefont{Berg}},
  \bibinfo{author}{\bibfnamefont{M.~S.} \bibnamefont{Rudner}},
  \bibinfo{author}{\bibfnamefont{G.}~\bibnamefont{Refael}}, \bibnamefont{and}
  \bibinfo{author}{\bibfnamefont{N.~H.} \bibnamefont{Lindner}},
  \bibinfo{journal}{Phys. Rev. X} \textbf{\bibinfo{volume}{6}},
  \bibinfo{pages}{021013} (\bibinfo{year}{2016}).

\bibitem[{\citenamefont{Thouless}(1983{\natexlab{a}})}]{Thouless1983Pump}
\bibinfo{author}{\bibfnamefont{D.~J.} \bibnamefont{Thouless}},
  \bibinfo{journal}{Phys. Rev. B.} \textbf{\bibinfo{volume}{27}},
  \bibinfo{pages}{6083} (\bibinfo{year}{1983}{\natexlab{a}}).

\bibitem[{\citenamefont{Kolodrubetz et~al.}(2018)\citenamefont{Kolodrubetz,
  Nathan, Gazit, Morimoto, and Moore}}]{EnergyPumpPaper}
\bibinfo{author}{\bibfnamefont{M.~H.} \bibnamefont{Kolodrubetz}},
  \bibinfo{author}{\bibfnamefont{F.}~\bibnamefont{Nathan}},
  \bibinfo{author}{\bibfnamefont{S.}~\bibnamefont{Gazit}},
  \bibinfo{author}{\bibfnamefont{T.}~\bibnamefont{Morimoto}}, \bibnamefont{and}
  \bibinfo{author}{\bibfnamefont{J.~E.} \bibnamefont{Moore}},
  \bibinfo{journal}{Phys. Rev. Lett.} \textbf{\bibinfo{volume}{120}},
  \bibinfo{pages}{150601} (\bibinfo{year}{2018}),
  \urlprefix\url{https://link.aps.org/doi/10.1103/PhysRevLett.120.150601}.

\bibitem[{\citenamefont{Peng and Refael}(2018)}]{Peng2018}
\bibinfo{author}{\bibfnamefont{Y.}~\bibnamefont{Peng}} \bibnamefont{and}
  \bibinfo{author}{\bibfnamefont{G.}~\bibnamefont{Refael}},
  \bibinfo{journal}{ArXiv e-prints}  (\bibinfo{year}{2018}),
  \eprint{1801.05811}, \urlprefix\url{https://arxiv.org/pdf/1801.05811}.

\bibitem[{\citenamefont{Weinberg et~al.}(2017)\citenamefont{Weinberg, Bukov,
  D'Alessio, Polkovnikov, Vajna, and Kolodrubetz}}]{Weinberg2017}
\bibinfo{author}{\bibfnamefont{P.}~\bibnamefont{Weinberg}},
  \bibinfo{author}{\bibfnamefont{M.}~\bibnamefont{Bukov}},
  \bibinfo{author}{\bibfnamefont{L.}~\bibnamefont{D'Alessio}},
  \bibinfo{author}{\bibfnamefont{A.}~\bibnamefont{Polkovnikov}},
  \bibinfo{author}{\bibfnamefont{S.}~\bibnamefont{Vajna}}, \bibnamefont{and}
  \bibinfo{author}{\bibfnamefont{M.}~\bibnamefont{Kolodrubetz}},
  \bibinfo{journal}{Physics Reports} \textbf{\bibinfo{volume}{688}},
  \bibinfo{pages}{1} (\bibinfo{year}{2017}), ISSN \bibinfo{issn}{0370-1573}.

\bibitem[{\citenamefont{Crowley et~al.}(2018)\citenamefont{Crowley, Martin, and
  Chandran}}]{Crowley2018}
\bibinfo{author}{\bibfnamefont{P.~J.~D.} \bibnamefont{Crowley}},
  \bibinfo{author}{\bibfnamefont{I.}~\bibnamefont{Martin}}, \bibnamefont{and}
  \bibinfo{author}{\bibfnamefont{A.}~\bibnamefont{Chandran}},
  \bibinfo{journal}{arXiv:1708.05023}  (\bibinfo{year}{2018}).

\bibitem[{\citenamefont{Haldane and Raghu}(2008)}]{Haldane2008}
\bibinfo{author}{\bibfnamefont{F.~D.~M.} \bibnamefont{Haldane}}
  \bibnamefont{and} \bibinfo{author}{\bibfnamefont{S.}~\bibnamefont{Raghu}},
  \bibinfo{journal}{Physical Review Letters} \textbf{\bibinfo{volume}{100}}
  (\bibinfo{year}{2008}).

\bibitem[{\citenamefont{Khanikaev et~al.}(2012)\citenamefont{Khanikaev,
  Hossein~Mousavi, Tse, Kargarian, MacDonald, and Shvets}}]{Khanikaev2012}
\bibinfo{author}{\bibfnamefont{A.~B.} \bibnamefont{Khanikaev}},
  \bibinfo{author}{\bibfnamefont{S.}~\bibnamefont{Hossein~Mousavi}},
  \bibinfo{author}{\bibfnamefont{W.-K.} \bibnamefont{Tse}},
  \bibinfo{author}{\bibfnamefont{M.}~\bibnamefont{Kargarian}},
  \bibinfo{author}{\bibfnamefont{A.~H.} \bibnamefont{MacDonald}},
  \bibnamefont{and} \bibinfo{author}{\bibfnamefont{G.}~\bibnamefont{Shvets}},
  \bibinfo{journal}{Nature Materials} \textbf{\bibinfo{volume}{12}},
  \bibinfo{pages}{233 EP } (\bibinfo{year}{2012}),
  \urlprefix\url{http://dx.doi.org/10.1038/nmat3520}.

\bibitem[{\citenamefont{Rechtsman et~al.}(2013)\citenamefont{Rechtsman, Zeuner,
  Plotnik, Lumer, Podolsky, Dreisow, Nolte, Segev, and
  Szameit}}]{Rechtsman2013}
\bibinfo{author}{\bibfnamefont{M.~C.} \bibnamefont{Rechtsman}},
  \bibinfo{author}{\bibfnamefont{J.~M.} \bibnamefont{Zeuner}},
  \bibinfo{author}{\bibfnamefont{Y.}~\bibnamefont{Plotnik}},
  \bibinfo{author}{\bibfnamefont{Y.}~\bibnamefont{Lumer}},
  \bibinfo{author}{\bibfnamefont{D.}~\bibnamefont{Podolsky}},
  \bibinfo{author}{\bibfnamefont{F.}~\bibnamefont{Dreisow}},
  \bibinfo{author}{\bibfnamefont{S.}~\bibnamefont{Nolte}},
  \bibinfo{author}{\bibfnamefont{M.}~\bibnamefont{Segev}}, \bibnamefont{and}
  \bibinfo{author}{\bibfnamefont{A.}~\bibnamefont{Szameit}},
  \bibinfo{journal}{Nature} \textbf{\bibinfo{volume}{496}},
  \bibinfo{pages}{196} (\bibinfo{year}{2013}).

\bibitem[{\citenamefont{Gao et~al.}(2016)\citenamefont{Gao, Gao, Shi, Yang,
  Lin, Xu, Joannopoulos, Solja{\v c}i{\'c}, Chen, Lu et~al.}}]{Gao2016}
\bibinfo{author}{\bibfnamefont{F.}~\bibnamefont{Gao}},
  \bibinfo{author}{\bibfnamefont{Z.}~\bibnamefont{Gao}},
  \bibinfo{author}{\bibfnamefont{X.}~\bibnamefont{Shi}},
  \bibinfo{author}{\bibfnamefont{Z.}~\bibnamefont{Yang}},
  \bibinfo{author}{\bibfnamefont{X.}~\bibnamefont{Lin}},
  \bibinfo{author}{\bibfnamefont{H.}~\bibnamefont{Xu}},
  \bibinfo{author}{\bibfnamefont{J.~D.} \bibnamefont{Joannopoulos}},
  \bibinfo{author}{\bibfnamefont{M.}~\bibnamefont{Solja{\v c}i{\'c}}},
  \bibinfo{author}{\bibfnamefont{H.}~\bibnamefont{Chen}},
  \bibinfo{author}{\bibfnamefont{L.}~\bibnamefont{Lu}}, \bibnamefont{et~al.},
  \bibinfo{journal}{Nature Communications} \textbf{\bibinfo{volume}{7}},
  \bibinfo{pages}{11619} (\bibinfo{year}{2016}), ISSN
  \bibinfo{issn}{2041-1723},
  \urlprefix\url{http://dx.doi.org/10.1038/ncomms11619}.

\bibitem[{\citenamefont{Mukherjee et~al.}(2017)\citenamefont{Mukherjee,
  Spracklen, Valiente, Andersson, {\"O}hberg, Goldman, and
  Thomson}}]{Mukherjee2017}
\bibinfo{author}{\bibfnamefont{S.}~\bibnamefont{Mukherjee}},
  \bibinfo{author}{\bibfnamefont{A.}~\bibnamefont{Spracklen}},
  \bibinfo{author}{\bibfnamefont{M.}~\bibnamefont{Valiente}},
  \bibinfo{author}{\bibfnamefont{E.}~\bibnamefont{Andersson}},
  \bibinfo{author}{\bibfnamefont{P.}~\bibnamefont{{\"O}hberg}},
  \bibinfo{author}{\bibfnamefont{N.}~\bibnamefont{Goldman}}, \bibnamefont{and}
  \bibinfo{author}{\bibfnamefont{R.~R.} \bibnamefont{Thomson}},
  \bibinfo{journal}{Nature Communications} \textbf{\bibinfo{volume}{8}},
  \bibinfo{pages}{13918 EP } (\bibinfo{year}{2017}),
  \urlprefix\url{http://dx.doi.org/10.1038/ncomms13918}.

\bibitem[{\citenamefont{Maczewsky et~al.}(2017)\citenamefont{Maczewsky, Zeuner,
  Nolte, and Szameit}}]{Maczewsky2017}
\bibinfo{author}{\bibfnamefont{L.~J.} \bibnamefont{Maczewsky}},
  \bibinfo{author}{\bibfnamefont{J.~M.} \bibnamefont{Zeuner}},
  \bibinfo{author}{\bibfnamefont{S.}~\bibnamefont{Nolte}}, \bibnamefont{and}
  \bibinfo{author}{\bibfnamefont{A.}~\bibnamefont{Szameit}},
  \bibinfo{journal}{Nature Communications} \textbf{\bibinfo{volume}{8}},
  \bibinfo{pages}{13756 EP } (\bibinfo{year}{2017}),
  \urlprefix\url{http://dx.doi.org/10.1038/ncomms13756}.

\bibitem[{\citenamefont{Barik et~al.}(2018)\citenamefont{Barik, Karasahin,
  Flower, Cai, Miyake, DeGottardi, Hafezi, and Waks}}]{Barik2018}
\bibinfo{author}{\bibfnamefont{S.}~\bibnamefont{Barik}},
  \bibinfo{author}{\bibfnamefont{A.}~\bibnamefont{Karasahin}},
  \bibinfo{author}{\bibfnamefont{C.}~\bibnamefont{Flower}},
  \bibinfo{author}{\bibfnamefont{T.}~\bibnamefont{Cai}},
  \bibinfo{author}{\bibfnamefont{H.}~\bibnamefont{Miyake}},
  \bibinfo{author}{\bibfnamefont{W.}~\bibnamefont{DeGottardi}},
  \bibinfo{author}{\bibfnamefont{M.}~\bibnamefont{Hafezi}}, \bibnamefont{and}
  \bibinfo{author}{\bibfnamefont{E.}~\bibnamefont{Waks}},
  \bibinfo{journal}{Science} \textbf{\bibinfo{volume}{359}},
  \bibinfo{pages}{666} (\bibinfo{year}{2018}),
  \urlprefix\url{http://science.sciencemag.org/content/359/6376/666.abstract}.

\bibitem[{\citenamefont{Kane and Lubensky}(2013)}]{Kane2013}
\bibinfo{author}{\bibfnamefont{C.~L.} \bibnamefont{Kane}} \bibnamefont{and}
  \bibinfo{author}{\bibfnamefont{T.~C.} \bibnamefont{Lubensky}},
  \bibinfo{journal}{Nature Physics} \textbf{\bibinfo{volume}{10}},
  \bibinfo{pages}{39 EP } (\bibinfo{year}{2013}),
  \urlprefix\url{http://dx.doi.org/10.1038/nphys2835}.

\bibitem[{\citenamefont{Paulose et~al.}(2015)\citenamefont{Paulose, Chen, and
  Vitelli}}]{Paulose2015}
\bibinfo{author}{\bibfnamefont{J.}~\bibnamefont{Paulose}},
  \bibinfo{author}{\bibfnamefont{B.~G.-g.} \bibnamefont{Chen}},
  \bibnamefont{and} \bibinfo{author}{\bibfnamefont{V.}~\bibnamefont{Vitelli}},
  \bibinfo{journal}{Nature Physics} \textbf{\bibinfo{volume}{11}},
  \bibinfo{pages}{153 EP } (\bibinfo{year}{2015}),
  \urlprefix\url{http://dx.doi.org/10.1038/nphys3185}.

\bibitem[{\citenamefont{S{\"u}sstrunk and Huber}(2015)}]{Susstrunk2015}
\bibinfo{author}{\bibfnamefont{R.}~\bibnamefont{S{\"u}sstrunk}}
  \bibnamefont{and} \bibinfo{author}{\bibfnamefont{S.~D.} \bibnamefont{Huber}},
  \bibinfo{journal}{Science} \textbf{\bibinfo{volume}{349}},
  \bibinfo{pages}{47} (\bibinfo{year}{2015}),
  \urlprefix\url{http://science.sciencemag.org/content/349/6243/47.abstract}.

\bibitem[{\citenamefont{Peng et~al.}(2016)\citenamefont{Peng, Qin, Zhao, Shen,
  Xu, Bao, Jia, and Zhu}}]{Peng2016}
\bibinfo{author}{\bibfnamefont{Y.-G.} \bibnamefont{Peng}},
  \bibinfo{author}{\bibfnamefont{C.-Z.} \bibnamefont{Qin}},
  \bibinfo{author}{\bibfnamefont{D.-G.} \bibnamefont{Zhao}},
  \bibinfo{author}{\bibfnamefont{Y.-X.} \bibnamefont{Shen}},
  \bibinfo{author}{\bibfnamefont{X.-Y.} \bibnamefont{Xu}},
  \bibinfo{author}{\bibfnamefont{M.}~\bibnamefont{Bao}},
  \bibinfo{author}{\bibfnamefont{H.}~\bibnamefont{Jia}}, \bibnamefont{and}
  \bibinfo{author}{\bibfnamefont{X.-F.} \bibnamefont{Zhu}},
  \bibinfo{journal}{Nature Communications} \textbf{\bibinfo{volume}{7}},
  \bibinfo{pages}{13368} (\bibinfo{year}{2016}), ISSN
  \bibinfo{issn}{2041-1723},
  \urlprefix\url{http://dx.doi.org/10.1038/ncomms13368}.

\bibitem[{\citenamefont{Cardano et~al.}(2017)\citenamefont{Cardano, D'Errico,
  Dauphin, Maffei, Piccirillo, de~Lisio, De~Filippis, Cataudella, Santamato,
  Marrucci et~al.}}]{Cardano2017}
\bibinfo{author}{\bibfnamefont{F.}~\bibnamefont{Cardano}},
  \bibinfo{author}{\bibfnamefont{A.}~\bibnamefont{D'Errico}},
  \bibinfo{author}{\bibfnamefont{A.}~\bibnamefont{Dauphin}},
  \bibinfo{author}{\bibfnamefont{M.}~\bibnamefont{Maffei}},
  \bibinfo{author}{\bibfnamefont{B.}~\bibnamefont{Piccirillo}},
  \bibinfo{author}{\bibfnamefont{C.}~\bibnamefont{de~Lisio}},
  \bibinfo{author}{\bibfnamefont{G.}~\bibnamefont{De~Filippis}},
  \bibinfo{author}{\bibfnamefont{V.}~\bibnamefont{Cataudella}},
  \bibinfo{author}{\bibfnamefont{E.}~\bibnamefont{Santamato}},
  \bibinfo{author}{\bibfnamefont{L.}~\bibnamefont{Marrucci}},
  \bibnamefont{et~al.}, \bibinfo{journal}{Nature Communications}
  \textbf{\bibinfo{volume}{8}}, \bibinfo{pages}{15516} (\bibinfo{year}{2017}),
  ISSN \bibinfo{issn}{2041-1723},
  \urlprefix\url{http://dx.doi.org/10.1038/ncomms15516}.

\bibitem[{\citenamefont{Martin et~al.}(2017)\citenamefont{Martin, Refael, and
  Halperin}}]{Martin2017}
\bibinfo{author}{\bibfnamefont{I.}~\bibnamefont{Martin}},
  \bibinfo{author}{\bibfnamefont{G.}~\bibnamefont{Refael}}, \bibnamefont{and}
  \bibinfo{author}{\bibfnamefont{B.}~\bibnamefont{Halperin}},
  \bibinfo{journal}{Phys. Rev. X} \textbf{\bibinfo{volume}{7}},
  \bibinfo{pages}{041008} (\bibinfo{year}{2017}), \eprint{1612.02143},
  \urlprefix\url{https://arxiv.org/pdf/1612.02143}.

\bibitem[{\citenamefont{Nathan}()}]{MarkovLindblad}
\bibinfo{author}{\bibfnamefont{F.}~\bibnamefont{Nathan}}, \bibinfo{note}{to
  appear shortly}.

\bibitem[{\citenamefont{M{\o}lmer et~al.}(1993)\citenamefont{M{\o}lmer, Castin,
  and Dalibard}}]{Molmer1993}
\bibinfo{author}{\bibfnamefont{K.}~\bibnamefont{M{\o}lmer}},
  \bibinfo{author}{\bibfnamefont{Y.}~\bibnamefont{Castin}}, \bibnamefont{and}
  \bibinfo{author}{\bibfnamefont{J.}~\bibnamefont{Dalibard}},
  \bibinfo{journal}{Journal of the Optical Society of America B}
  \textbf{\bibinfo{volume}{10}}, \bibinfo{pages}{524} (\bibinfo{year}{1993}).

\bibitem[{\citenamefont{Thouless}(1983{\natexlab{b}})}]{ThoulessPump}
\bibinfo{author}{\bibfnamefont{D.~J.} \bibnamefont{Thouless}},
  \bibinfo{journal}{Phys. Rev. B.} \textbf{\bibinfo{volume}{27}},
  \bibinfo{pages}{6083} (\bibinfo{year}{1983}{\natexlab{b}}).

\bibitem[{\citenamefont{Gardiner and Zoller}(2004)}]{GardinerZoller}
\bibinfo{author}{\bibfnamefont{C.}~\bibnamefont{Gardiner}} \bibnamefont{and}
  \bibinfo{author}{\bibfnamefont{P.}~\bibnamefont{Zoller}},
  \emph{\bibinfo{title}{Quantum Noise}} (\bibinfo{publisher}{Springer-Verlag
  Berlin Heidelberg}, \bibinfo{year}{2004}).

\bibitem[{\citenamefont{Weiss}(1999)}]{QuantumDissipativeSystems}
\bibinfo{author}{\bibfnamefont{U.}~\bibnamefont{Weiss}},
  \emph{\bibinfo{title}{Quantum Dissipative Systems}}
  (\bibinfo{publisher}{WORLD SCIENTIFIC}, \bibinfo{year}{1999}), ISBN
  \bibinfo{isbn}{9789812817877},
  \urlprefix\url{http://dx.doi.org/10.1142/9789812817877_fmatter}.

\bibitem[{FN:({\natexlab{a}})}]{FN:FloquetStates}
\bibinfo{note}{Here the Floquet states are the unique sets of states whose
  evolution is given by a linearly increasing phase times a time-periodic
  state: $U(t)|\left.\! \phi_a(0)\right> = e^{-i\varepsilon _a t}|\left.\!
  \phi_a(t)\right>$, where $|\left.\! \phi_a(t)\right> =
  |\left.\!\phi_a(t+T)\right>$. The Floquet theorem dictates that a complete
  orthonormal basis of states with this property always exists for periodically
  driven systems.}

\bibitem[{FN:({\natexlab{b}})}]{FN:PoissonBracket}
\bibinfo{note}{To see this, note that the Hamiltonian of an isolated cavity
  mode with frequency $\omega$ is given by $H_{\rm cav} = B_c^2 V/\mu$, while
  the phase of the mode increases at the linear rate $\omega$: $\partial _t
  \phi = \omega$. Since $\partial _t \phi = \{\phi,H_{\rm cav}\}$, we must have
  $\{\phi,B_c\} = \frac{1}{2B_c}\frac{\omega \mu}{V}$.}

\bibitem[{\citenamefont{Nathan et~al.}()\citenamefont{Nathan, Refael, and
  Martin}}]{FollowUpWork}
\bibinfo{author}{\bibfnamefont{F.}~\bibnamefont{Nathan}},
  \bibinfo{author}{\bibfnamefont{G.}~\bibnamefont{Refael}}, \bibnamefont{and}
  \bibinfo{author}{\bibfnamefont{I.}~\bibnamefont{Martin}}, \bibinfo{note}{to
  appear shortly}.

\bibitem[{\citenamefont{Perot and Fabry}(1899)}]{Perot1899}
\bibinfo{author}{\bibfnamefont{A.}~\bibnamefont{Perot}} \bibnamefont{and}
  \bibinfo{author}{\bibfnamefont{C.}~\bibnamefont{Fabry}},
  \bibinfo{journal}{Astrophysical Journal} \textbf{\bibinfo{volume}{9}},
  \bibinfo{pages}{87} (\bibinfo{year}{1899}).

\bibitem[{\citenamefont{Hernandez}(1986)}]{Hernandez1986}
\bibinfo{author}{\bibfnamefont{G.}~\bibnamefont{Hernandez}},
  \emph{\bibinfo{title}{Fabry--P{\'e}rot Interferometers}}
  (\bibinfo{publisher}{Cambridge University Press},
  \bibinfo{address}{Cambridge}, \bibinfo{year}{1986}), ISBN
  \bibinfo{isbn}{0-521-32238-3}.

\bibitem[{\citenamefont{Ismail et~al.}(2016)\citenamefont{Ismail, Kores,
  Geskus, and Pollnau}}]{Ismail2016FabryPerot}
\bibinfo{author}{\bibfnamefont{N.}~\bibnamefont{Ismail}},
  \bibinfo{author}{\bibfnamefont{C.~C.} \bibnamefont{Kores}},
  \bibinfo{author}{\bibfnamefont{D.}~\bibnamefont{Geskus}}, \bibnamefont{and}
  \bibinfo{author}{\bibfnamefont{M.}~\bibnamefont{Pollnau}},
  \bibinfo{journal}{Optics Express} \textbf{\bibinfo{volume}{24}},
  \bibinfo{pages}{16366} (\bibinfo{year}{2016}).

\bibitem[{\citenamefont{Born and Wolf}(1999)}]{Born1999}
\bibinfo{author}{\bibfnamefont{M.}~\bibnamefont{Born}} \bibnamefont{and}
  \bibinfo{author}{\bibfnamefont{E.}~\bibnamefont{Wolf}},
  \emph{\bibinfo{title}{Principles of Optics, Cambridge University Press}}
  (\bibinfo{publisher}{Cambridge University Press}, \bibinfo{year}{1999}).

\bibitem[{\citenamefont{Aurand et~al.}(2011)\citenamefont{Aurand, Kuschel,
  R{\"o}del, Heyer, Wunderlich, J{\"a}ckel, Kaluza, Paulus, and
  K{\"u}hl}}]{Aurand2011}
\bibinfo{author}{\bibfnamefont{B.}~\bibnamefont{Aurand}},
  \bibinfo{author}{\bibfnamefont{S.}~\bibnamefont{Kuschel}},
  \bibinfo{author}{\bibfnamefont{C.}~\bibnamefont{R{\"o}del}},
  \bibinfo{author}{\bibfnamefont{M.}~\bibnamefont{Heyer}},
  \bibinfo{author}{\bibfnamefont{F.}~\bibnamefont{Wunderlich}},
  \bibinfo{author}{\bibfnamefont{O.}~\bibnamefont{J{\"a}ckel}},
  \bibinfo{author}{\bibfnamefont{M.~C.} \bibnamefont{Kaluza}},
  \bibinfo{author}{\bibfnamefont{G.~G.} \bibnamefont{Paulus}},
  \bibnamefont{and} \bibinfo{author}{\bibfnamefont{T.}~\bibnamefont{K{\"u}hl}},
  \bibinfo{journal}{Optics Express} \textbf{\bibinfo{volume}{19}},
  \bibinfo{pages}{17151} (\bibinfo{year}{2011}), ISSN
  \bibinfo{issn}{1094-4087},
  \urlprefix\url{http://dx.doi.org/10.1364/OE.19.017151}.

\bibitem[{FN:({\natexlab{c}})}]{FN:TimeReversedPartner}
\bibinfo{note}{The slower decay of mode $-1$ is due to the fact that this
  mode's time-reversed partner, mode $1$, is excited. Thus $\vec L(t)$ thus has
  a large Fourier component at frequency $\pm \omega_1$. Since $\omega_{-1}
  \approx - \omega_1$, the coefficient $\Phi_{-1}$ should take a relatively
  large value.}

\bibitem[{\citenamefont{Tyryshkin et~al.}(2003)\citenamefont{Tyryshkin, Lyon,
  Astashkin, and Raitsimring}}]{Tyryshkin2003}
\bibinfo{author}{\bibfnamefont{A.~M.} \bibnamefont{Tyryshkin}},
  \bibinfo{author}{\bibfnamefont{S.~A.} \bibnamefont{Lyon}},
  \bibinfo{author}{\bibfnamefont{A.~V.} \bibnamefont{Astashkin}},
  \bibnamefont{and} \bibinfo{author}{\bibfnamefont{A.~M.}
  \bibnamefont{Raitsimring}}, \bibinfo{journal}{Physical Review B}
  \textbf{\bibinfo{volume}{68}} (\bibinfo{year}{2003}).

\bibitem[{\citenamefont{Singh et~al.}(2012)\citenamefont{Singh, Sharma, Lin,
  Hasan, Prasad, and Bansil}}]{Singh2012}
\bibinfo{author}{\bibfnamefont{B.}~\bibnamefont{Singh}},
  \bibinfo{author}{\bibfnamefont{A.}~\bibnamefont{Sharma}},
  \bibinfo{author}{\bibfnamefont{H.}~\bibnamefont{Lin}},
  \bibinfo{author}{\bibfnamefont{M.~Z.} \bibnamefont{Hasan}},
  \bibinfo{author}{\bibfnamefont{R.}~\bibnamefont{Prasad}}, \bibnamefont{and}
  \bibinfo{author}{\bibfnamefont{A.}~\bibnamefont{Bansil}},
  \bibinfo{journal}{Physical Review B} \textbf{\bibinfo{volume}{86}}
  (\bibinfo{year}{2012}).

\bibitem[{\citenamefont{Wan et~al.}(2011)\citenamefont{Wan, Turner, Vishwanath,
  and Savrasov}}]{Wan2011}
\bibinfo{author}{\bibfnamefont{X.}~\bibnamefont{Wan}},
  \bibinfo{author}{\bibfnamefont{A.~M.} \bibnamefont{Turner}},
  \bibinfo{author}{\bibfnamefont{A.}~\bibnamefont{Vishwanath}},
  \bibnamefont{and} \bibinfo{author}{\bibfnamefont{S.~Y.}
  \bibnamefont{Savrasov}}, \bibinfo{journal}{Phys. Rev. B}
  \textbf{\bibinfo{volume}{83}}, \bibinfo{pages}{205101}
  (\bibinfo{year}{2011}).

\bibitem[{\citenamefont{Burkov and Balents}(2011)}]{Burkow2011a}
\bibinfo{author}{\bibfnamefont{A.~A.} \bibnamefont{Burkov}} \bibnamefont{and}
  \bibinfo{author}{\bibfnamefont{L.}~\bibnamefont{Balents}},
  \bibinfo{journal}{Physical Review Letters} \textbf{\bibinfo{volume}{107}}
  (\bibinfo{year}{2011}).

\bibitem[{\citenamefont{Burkov et~al.}(2011)\citenamefont{Burkov, Hook, and
  Balents}}]{Burkov2011b}
\bibinfo{author}{\bibfnamefont{A.~A.} \bibnamefont{Burkov}},
  \bibinfo{author}{\bibfnamefont{M.~D.} \bibnamefont{Hook}}, \bibnamefont{and}
  \bibinfo{author}{\bibfnamefont{L.}~\bibnamefont{Balents}},
  \bibinfo{journal}{Physical Review B} \textbf{\bibinfo{volume}{84}}
  (\bibinfo{year}{2011}).

\end{thebibliography}

\end{document}